\documentclass[twocolumn]{aastex631}

\usepackage{hyperref}
\usepackage{subfigure}
\usepackage{graphicx}
\usepackage[flushleft]{threeparttable}

\begin{document}

\title{Accretion Geometry of GX~339--4 in the Hard State: AstroSat View}

\correspondingauthor{Swadesh Chand}
\email{swadesh.chand@iucaa.in}

\author{Swadesh Chand}
\affiliation{Inter-University Center for Astronomy and Astrophysics, Pune, Maharashtra-411007, India}

\author{Gulab C. Dewangan}
\affiliation{Inter-University Center for Astronomy and Astrophysics, Pune, Maharashtra-411007, India}

\author{Andrzej A. Zdziarski}
\affiliation{Nicolaus Copernicus Astronomical Center, Polish Academy of Sciences, Bartycka 18, PL-00-716 Warszawa, Poland}

\author{Dipankar Bhattacharya}
\affiliation{Inter-University Center for Astronomy and Astrophysics, Pune, Maharashtra-411007, India}
\affiliation{Department of Physics, Ashoka University Rai, Sonipat, Haryana-131029, India}

\author{N. P. S. Mithun}
\affiliation{Physical Research Laboratory Thaltej, Ahmedabad, Gujarat 380009, India}

\author{Santosh V. Vadawale}
\affiliation{Physical Research Laboratory Thaltej, Ahmedabad, Gujarat 380009, India}







\begin{abstract}
We perform broadband ($0.7-100$ keV) spectral analysis of five hard state observations of the low-mass back hole X-ray binary GX~339--4 taken by AstroSat during the rising phase of three outbursts from $2019$ to $2022$. We find that the outburst in 2021 was the only successful/full outburst, while the source was unable to make transition to the soft state during the other two outbursts in 2019 and 2022. Our spectral analysis employs two different
model combinations, requiring two separate Comptonizing regions and their associated reflection components, and soft X-ray excess emission. The harder Comptonizing component dominates the overall bolometric luminosity, while the softer one remains relatively weak. Our spectral fits indicate that the disk evolves with the source luminosity,
where the inner disk radius decreases with increasing luminosity. However, the disk remains substantially truncated throughout all the observations at the source luminosity of $\sim2-8\%\times$ of the Eddington luminosity. We note that our assumption of the soft X-ray excess emission as disk blackbody may not be realistic, and this
kind of soft excess may arise due the non-homogeneity in the disk/corona geometry. Our temporal analysis deriving  the power density spectra suggests that the 
break frequency increases with the source luminosity. Furthermore, our analysis demonstrates a consistency between the inner disk radii estimated from break frequency of the power density spectra and those obtained from the reflection modelling, supporting the truncated disk geometry in the hard state.
\end{abstract}



\section{Introduction} \label{sec:intro}

Most of the low-mass black hole X-ray binaries (BHXRBs) show occasional outbursts due to their transient nature. These systems spend most of their lifetime in quiescence and get powered by the mass accretion process from the companion star onto the stellar-mass black hole (BH) during outbursts. A sudden change in the mass accretion rate can cause the system to make a transition from the quiescence to the outburst phase, which can last up to several years \citep{2001A&A...373..251D, 2009MNRAS.400.1337D}. 

The evolution of an outburst brings about specific changes in the spectral and timing properties of these black hole transients (BHTs). Following quiescence and at the early phase of an outburst, BHTs are observed in the low/hard state (LHS), wherein a power-law component and significant aperiodic variability ($\sim30\%$) dominate. As the mass accretion rate and consequently the luminosity increase, these sources eventually transition into the high/soft state (HSS), where the contribution from the accretion disk becomes dominant and the aperiodic variability diminishes to only a few percent. Additionally, there exist two intermediate states known as the hard intermediate (HIMS) and the soft-intermediate (SIMS) states, through which the BHTs progress during the transition from the LHS to HSS. Both intermediate states exhibit substantial contributions from both the disk and the power-law components; however, they are distinguishable from each other based on their distinct timing properties. The different spectral states of BHTs during a full outburst can be identified through a q-shaped diagram known as the hardness-intensity diagram \citep[HID;][]{2005A&A...440..207B, 2005Ap&SS.300..107H}. In addition to the comprehensive outburst evolution, many instances have been observed where the BHTs were unable to reach the soft state and instead remained in the LHS throughout the outburst. Such outbursts are typically termed as `hard-only' or `failed' outbursts, and they typically have a comparatively shorter duration and lower luminosity than full outbursts.

In the HSS, it is very well established that the optically thick and geometrically thin accretion disk \citep{1973A&A....24..337S} extends all the way down to the inner most-stable circular orbit (ISCO), and the dominant emission from the accretion disk peaks at $\sim1$ keV \citep{2011MNRAS.411..337D, 2014MNRAS.442.1767P}. Study of this disk continuum can enable us to measure the inner disk radius and spin of the BH. On the other hand, the contribution from the disk emission is very low in the LHS, and the dominant X-ray power-law spectrum with a high energy cut-off at $\sim 100$~keV originates from the inverse Compton scattering of the soft seed photons from the disk by the hot electrons in an optically thin corona. Another crucial aspect of the broadband X-ray spectrum is the reflected component from the disk arising due to irradiation of coronal X-rays onto the disk. The reflection component  appears to be superimposed with the power-law continuum. The most prominent features of the reflection component are the iron $\mathrm{K_\alpha}$ line at $\sim6.4$ keV and reflection hump peaking at $\sim20-40$ keV. The iron $\mathrm{K_\alpha}$ line may appear to be broadened and skewed due to combined effects of Doppler shifts and general relativity if the reflecting material is close to the BH \citep{1989MNRAS.238..729F}. Modeling of the reflection spectrum has been proven to be one of the most efficient methods to measure the inner disk radius apart from the disk continuum approach. 

The actual geometry of the accretion disk in the LHS is still unclear, and the extent of the inner accretion disk in this state remains a topic of debate.
A number of studies suggest that the inner accretion disk in the LHS is most likely replaced by an optically thin and geometrically thick hot inner-accretion flow \citep{1997ApJ...489..865E}, thereby leaving the accretion disk to be truncated away from the ISCO. As the source undergoes  state transition along with the increase in mass accretion rate, the disk starts getting closer to the ISCO by penetrating the hot inner flow \citep{2015A&A...573A.120P, 2016MNRAS.458.2199B}. In addition, \cite{1996A&A...314..813L} suggested that a truncated accretion disk is necessary at the early phase of outburst as there is no inner disk during quiescence. However, the truncated accretion disk scenario in the LHS is questioned by several investigations that have detected a broad iron line in the LHS, indicating the accretion disk to be extended close to the ISCO \citep{2006ApJ...653..525M, 2008ApJ...679L.113M, 2008MNRAS.387.1489R}.  
But subsequent thorough investigations of these observations by \cite{2006MNRAS.367..659D}, \cite{2009ApJ...707L.109Y}, \cite{2010MNRAS.407.2287D} and \cite{2014MNRAS.437..316K} revealed the presence of a substantially truncated accretion disk along with narrow iron line, and suggested the earlier 
detection of broad lines were due to pile up in the detector. Apart from this, \cite{2019ApJ...885...48G} reported that the inner accretion disk reaches close to the ISCO during the bright hard state, when the luminosity becomes high, close to $1\%$ of the Eddington luminosity ($L_\mathrm{E}$). On the other hand, \cite{2020ApJ...896L..36Z} stated that the disk comes down to the ISCO only when the source reaches the HSS from the LHS. The dilemma over 
the extent of the inner accretion disk in the LHS persists, as two different measurements of the inner disk radius were estimated using
the same RXTE and Swift/XRT observations of the BHT XTE~J1752--223 in the long stable hard state during the 2009-2010 outburst, with \cite{2018ApJ...864...25G} finding the inner extent of the disk at ISCO and \cite{2021ApJ...906...69Z} estimating a truncated disk. In their work, 
\cite{2021ApJ...906...69Z} invoked double Comptonization components rather than a single one, and a series of different reflection model 
components. Subsequently, \cite{2022ApJ...935..118C} studied these observations again using high density reflection models with double Comptonization scenario, and found different constraints on the disk radius from reflection and disk components. Evidence of truncated accretion disk in the LHS of 
BHTs has also been reported by several other studies \citep{2015A&A...573A.120P, 2016MNRAS.458.2199B, 2020ApJ...893..142C}. In addition to this, \cite{2021ApJ...909L...9Z} studied the BHT MAXI~J1820+070 in the LHS during the rising phase of the 2018 outburst using two primary Comptonization regions and found that the accretion disk is significantly truncated away from the ISCO at the source luminosity of $\sim5-15 \%~L_\mathrm{E}$. 
Using AstroSat observations of MAXI J1820+070 during the 2018 outburst, \cite{2024arXiv240208237B} also performed a multi-wavelength spectral analysis and reported a substantially truncated accretion disk ($\gtrsim51~r_{\rm{g}}$) in the hard state with a double Comptonization scenario.

Study of the time variability in terms of the power density spectrum (PDS) can also provide crucial insights into the accretion flow mechanism. The PDSs in the LHS of BHTs usually show the dominant presence of band-limited noise (BLN) components, and sometimes narrow low-frequency quasi-periodic oscillations (QPOs) are also observed to be superimposed upon \citep{2005ApJ...629..403C, 2005Ap&SS.300..107H}. The BLN component remains flat up to a frequency, also known as the break frequency, above which a steepness in the slope is observed roughly as $\mathrm{P_\nu}\propto\nu^{-1}$ 
\citep{1995xrbi.nasa..252V, 2006csxs.book...39V, 2007A&ARv..15....1D, 2015A&A...573A.120P}. The timescale associated with the break 
frequency is found to be correlated with the viscous timescale of the system, and exhibits substantial evolution over the change of the spectral states of BHTs \citep{2007A&ARv..15....1D}. As the source luminosity increases and the power-law energy spectrum steepens from the dimmer to brighter LHS, the break frequency also increases, which may in turn be associated with the decrease in the truncation radius of the inner accretion disk \citep{1999A&A...352..182G, 2001MNRAS.321..759C, 2007A&ARv..15....1D}. Thus, the measurement of the break frequency appears to be a compelling alternative approach to estimate the inner disk radius. 

GX~339--4 is one of the most studied low-mass BHTs that has undergone multiple  outbursts since its discovery \citep{1973ApJ...184L..67M}. It has been a key source to probe the disk/corona geometry in the LHS. It is known to be located at a distance, $d$, of $\sim8-12$ kpc \citep{2004MNRAS.351..791Z, 2019MNRAS.488.1026Z}, and harbours a BH spinning
at or near the maximum \citep{2008MNRAS.387.1489R, 2008ApJ...679L.113M, 2015ApJ...806..262L, 2016ApJ...821L...6P}. The mass of the BH in GX~339--4 has not been estimated yet properly due to the position of the source near the Galactic disk region, which makes the dynamical measurement of the BH mass very complex. The measurement through the scaling of photon index and QPO frequency by \cite{2009ApJ...699..453S} suggests the BH mass, $M_{\rm{BH}}$, to be $12.3\pm1.4~M_\odot$. Furthermore, several latest estimations of BH mass of GX~339--4 are $9.0^{+1.6}_{-1.2}~M_\odot$ \citep{2016ApJ...821L...6P}, $2.3-9.5~M_\odot$ \citep{2017ApJ...846..132H}, $8.28-11.89~M_\odot$ \citep{2019AdSpR..63.1374S} and $\sim4-11~M_\odot$ \citep{2019MNRAS.488.1026Z}. The binary inclination angle, derived using the quiescence state observations by \cite{2017ApJ...846..132H} suggests it to be between $37^\circ$ and $78^\circ$. Moreover, \cite{2019MNRAS.488.1026Z} determined the inclination angle to be $\approx40^\circ-60^\circ$ from their evolutionary model constructed for the donor.

In this paper, we further explore the geometry of the accretion disk in the LHS of GX~339--4 using sets of observations  taken by India's first multi-wavelength satellite AstroSat during the 2019--2022 outbursts. Here, we have utilized all the three X-ray instruments onboard AstroSat, which allow us to study the broadband spectra in the $0.7-100$ keV band. We use two different model combinations including relativistic reflection components, and perform broadband spectral analysis in the framework of two Compotonization regions to understand the geometry of the source in the LHS. Throughout this work, we have assumed the BH mass $M_{\rm{BH}}=10M_\odot$, distance $d=8$ kpc \citep{2015ApJ...813...84G, 2019ApJ...885...48G}, and inclination $i=60^\circ$ \citep{2010MNRAS.407.2287D, 2010MNRAS.406.2206K, 2022MNRAS.510.4040H}. We have also discussed the reason for choosing the 
inclination angle in the analysis section. The remainder of the paper is organized as follows. We provide the details of the observations and data reduction in Section~\ref{sec:observation},  analysis and results in Section~\ref{sec:Analysis and Results}. Finally, Section~\ref{sec:discusiion} is devoted to discussion and concluding remarks.

\section{Observation and Data Reduction} \label{sec:observation}

\begin{figure}
    \centering
    \includegraphics[width=\columnwidth]{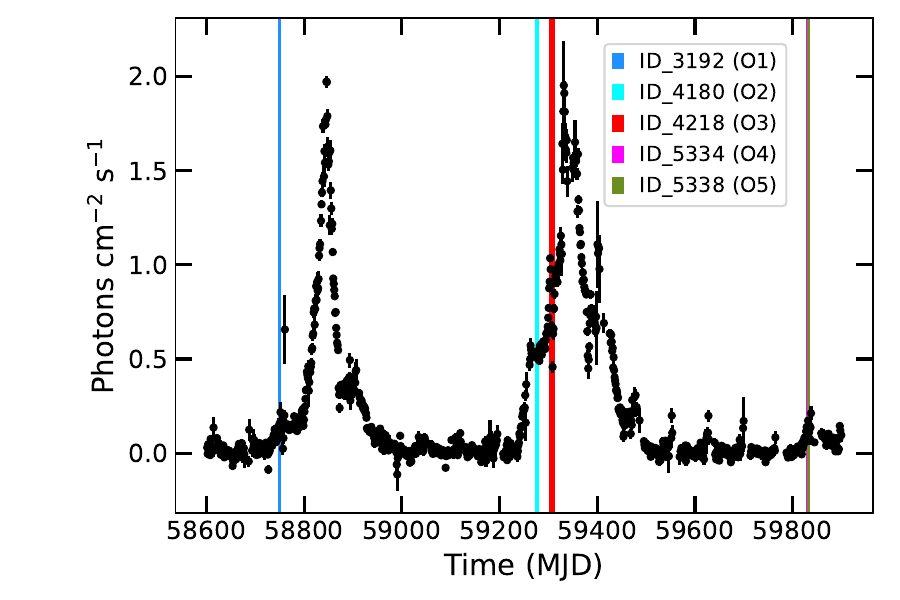}
    \caption{MAXI long-term lightcurve in the $2-20$ keV band covering the three outbursts of GX~339--4 in 2019, 2021 and 2022. The vertical
    lines indicate the positions of the AstroSat observations, where the associated IDs correspond to the last four digits of the observation 
    IDs listed in Table~\ref{tab:obs_ids}.}
    \label{fig:maxi_ltcrv}
\end{figure}

AstroSat has observed GX~339--4 many times during the period of its different outbursts. Here, we have analyzed five AstroSat observations of GX~339--4 during three different outbursts in 2019, 2021 and 2022. These observations, which have very good quality high energy spectral data up to 100 keV, were made during the rising phase of the outbursts, when the source was mostly in the hard state. The details of these observations are given in Table~\ref{tab:obs_ids}, and the timing of the AstroSat observations during these three outbursts are marked in a long term MAXI lightcurve of the source as shown in Figure~\ref{fig:maxi_ltcrv}. 

\subsection{Hardness-intensity Diagram}

\begin{figure*}
    \centering
    \includegraphics[width=0.48\linewidth]{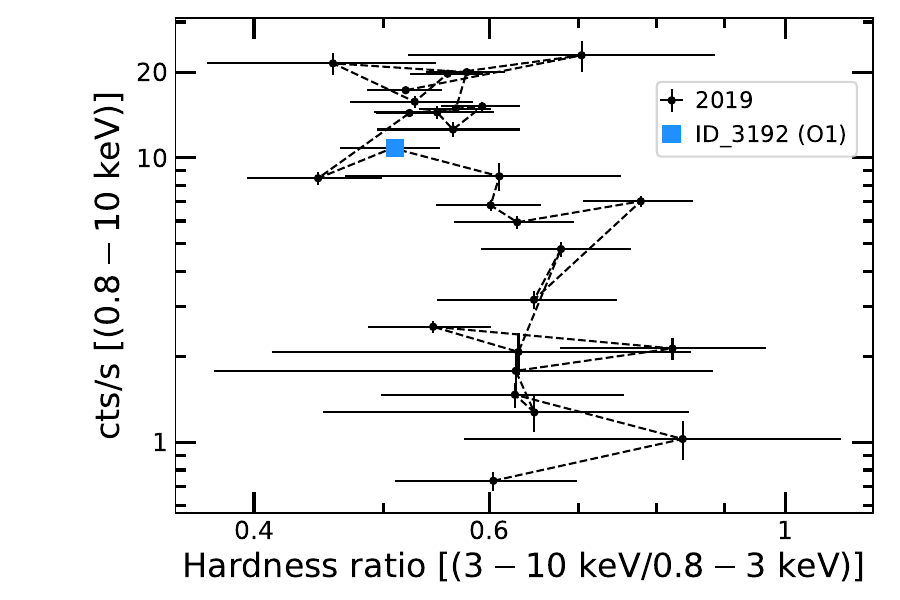}
    \includegraphics[width=0.48\linewidth]{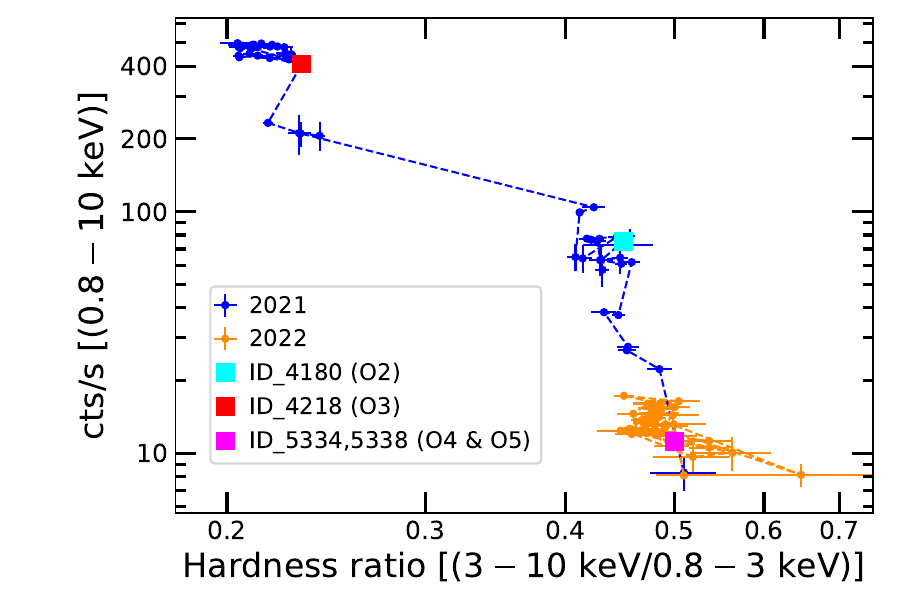}
    \caption{Hardness-intensity diagrams derived using Swift/XRT observations in 2019 (left panel), and 2021 and 2022 outbursts (right panel). The observations in 2019 were taken in the PC mode, whereas the observations in both the 2021 and 2022 in the WT mode. The associated IDs correspond to the last four digits of the observation IDs listed in Table~\ref{tab:obs_ids}.}
    \label{fig:xrt_hid}
\end{figure*}

In order to investigate the nature of the three outbursts, we derived the hardness-intensity diagam (HID) using the Swift/XRT observations taken during the 2019, 2021 and 2022 outbursts. We used the WT (windowed timing) mode data during the 2021 and 2022 outbursts and derived count rates in the $0.8-10$ keV, $0.8-3$ keV and $3-10$ keV bands using the online Swift/XRT data analysis tools provided by the Leicester Swift Data Centre\footnote{\url{https://www.swift.ac.uk/user_objects/}}. Since no WT mode data were available during the 2019 outburst, we used  the PC (photon counting) mode observations and derived count rates in the same energy bands mentioned for the WT mode data. The hardness ratio (HR) was then calculated by dividing the count rates in the $0.8-3$ keV band by the count rates in the $3-10$ keV band. Figure~\ref{fig:xrt_hid} shows the HIDs for the 2019 outburst (left panel), and the outbursts in 2021 and 2022 (right panel), as well as the positions of the five AstroSat observations across these outbursts. This figure shows a distinct spectral transition of the source during the 2021 outburst, whereas the source stayed in the LHS only throughout the 2019 and 2022 outbursts. The HIDs indicate that the 2021 outburst was a complete one, but the outbursts in 2019 and 2022 were `hard-only' or `failed' outbursts. 
Furthermore, during the third observation, a discernible decrease in the HR with the increase in the count rate suggests that the source
might have reached the HIMS during this observation. On the other hand, the source was in the LHS during the rest of the observations.

\subsection{AstroSat}

We downloaded the level2 data acquired with the Soft X-Ray Telescope  \citep[SXT;][]{2016SPIE.9905E..1ES, 2017JApA...38...29S} for all
the five observations directly from the AstroSat data archive{\footnote {\url{https://astrobrowse.issdc.gov.in/astro_archive/archive/Home.jsp}}}. Individual orbits from each observation were then merged using the SXT event merger tool {\footnote{\url{https://github.com/gulabd/SXTMerger.jl}}} to create an exposure corrected merged clean event file. We used this merged clean event file to extract the source spectrum from a circular region of radius $15\arcmin$ centered at the source position using the XSELECT V2.5b tool. The source count rates in each of the PC mode observation were found to be well below the threshold value for the pile-up i.e. $>40$ c/s, indicating that
none of these observations were affected by photon pile-up. Due to the large PSF (FWHM$\sim2\arcmin$) and half-power diameter ($\sim10\arcmin$) of the SXT, as well as the presence of the calibration sources at all the corners of the SXT CCD, it is not possible to extract the background spectrum from a source free region of the same CCD. Hence, we used the  blank sky background spectrum \texttt{`SkyBkg\_comb\_EL3p5\_Cl\_Rd16p0\_v01.pha'} provided by the SXT payload operation centre (POC). We used the response matrix file (RMF) \texttt{`sxt\_pc\_mat\_g0to12.rmf'} available at POC website. The SXT off-axis auxiliary response ﬁle (ARF), provided by the POC, was also modified as per the location of the source on the CCD using the sxtARFModule tool{\footnote{\url{https://www.tifr.res.in/~astrosat_sxt/dataanalysis.html}}}. Since, the shift in the gain of the onboard SXT instrument after the launch of the spacecraft has not been incorporated in the current version of the SXTPIPELINE, we corrected 
all the SXT spectral data for this gain shift{\footnote{\url{https://www.tifr.res.in/~astrosat_sxt/dataana_up/readme_sxt_arf_data_analysis.txt?file10=download}}} during spectral analysis using the gain command within XSPEC \citep[v12.12.1;][]{1996ASPC..101...17A}.

For the observations taken by the Large Area X-Ray Proportional Counter \citep[LAXPC;][]{2016SPIE.9905E..1DY, 2016ApJ...833...27Y, 2017JApA...38...30A, 2017ApJS..231...10A}, we processed the level1 data to level2 using the LAXPC software (Laxpcsoft){\footnote{\url{https://www.tifr.res.in/~astrosat_laxpc/LaxpcSoft.html}}}. In this present study, the data from the LAXPC20 detector are only used for spectral and temporal analysis. We note here that the background count rates were well below the source count 
rates up to $80$ keV for all the observations. Due to the low gain and response of the instrument, the data from the LAXPC10 detector was discarded. The data from the LAXPC30 detector are not available as this detector is no longer in operation.

We also used the data taken from the Cadmium Zinc Telluride Imager (CZTI) in order to study the hard X-ray spectra up to $100$ keV. Unlike the SXT and LAXPC instruments, the CZTI data are available in a single file after merging data from all the orbits together for each observation ID. We used these merged level1 data for each observation to produce clean event files using the new version of the CZTPIPELINE (v3.0){\footnote{\url{http://astrosat-ssc.iucaa.in/cztiData}}} and updated calibrations. This
version of the CZTPIPELINE incorporates an improved mask-weighting technique in its cztbindata module \citep[Mithun et al. in prep;][]{2024ApJ...960L...2C}. We used the clean event file in order to derive background subtracted spectra from each quadrant for each observation using the standard tasks within CZTPIPELINE{\footnote{\url{http://astrosat-ssc.iucaa.in/uploads/threadsPageNew_SXT.html}}}. Finally, we merged the spectra from all the four quadrant together to obtain a high signal-to-noise spectrum for each observation.

\begin{table*}
	\caption{List of AstroSat Observations of GX~339--4}
   \centering
   \begin{tabular}{lccccc}
      \hline
      \hline
    Obs No &  Obs ID & Obs. Start Date & Eff. Exp.(ks) & Count rate (c/s) \\
    &  & (yyyy-mm-dd) & SXT/LAXPC/CZTI & SXT/LAXPC/CZTI\\
     \hline
        O1 &  $9000003192$ & 2019-09-22 & $12/34/28.3$ & $4.6/109.1/4.6$  \\
     O2 &  $9000004180$ & 2021-02-13 & $10.6/29.9/25.3$ & $23.7/396.6/13.2$  \\
      O3 &  $9000004218$ & 2021-03-02 & $42.2/99.8/78.9$ & $29.3/466/14.2$ \\
    O4 &  $9000005334$ & 2022-09-07 & $17.3/38.9/36.9$ & $4.7/112/4.1$ \\
    O5 &  $9000005338$ & 2022-09-09 & $34/86.9/79$ & $1.4/120.5/4.5$ \\
     
     \hline
   \end{tabular}
   \label{tab:obs_ids}
   \begin{tablenotes}
    \item Note--The count rates are derived in the $0.7-6$ keV for SXT, $4-80$ keV for LAXPC20 and $25-100$ keV for CZTI.
    \end{tablenotes}
\end{table*}

\section{Analysis and Results} \label{sec:Analysis and Results}

\subsection{Broadband Spectral Analysis}

We performed broadband spectral analysis in the $(0.7-100)$ keV band using the SXT ($0.7-6$ keV), LAXPC20 ($4-80$ keV) and CZTI ($25-100$ keV) spectral data jointly within XSPEC. For the observation 5, we discarded the data below 0.8 keV to avoid the calibration uncertainties present at the lower energies. We quote the error bars on the best-fit spectral parameters at the $90\%$ confidence level unless otherwise specified. 
In order to apply the $\chi^2$ statistics, we used the optimal binning algorithm \citep{2016A&A...587A.151K} with a minimum of 20 counts per grouped bin for both the SXT and LAXPC20 spectral data. For the CZTI spectral data, the default binning criteria within the CZTPIPELINE(v3.0) has been used. We multiplied our models by a constant factor to account for the differences in the relative normalizations of the three instruments. We fixed this factor at 1 for the SXT data and kept free to vary for both the LAXPC20 and CZTI  data. As per the recommendation of the instrumentation teams, we used a systematic error of $3\%$ to the models during spectral modelling to account for the possible uncertainties in the calibration of different instruments{\footnote{\url{https://www.tifr.res.in/~astrosat_sxt/dataanalysis.html}}}.

We started our spectral analysis by using a 
\texttt{powerlaw} model modified by the Galactic absorption component \texttt{tbabs}, where the abundances are taken from \cite{2000ApJ...542..914W} and absorption cross sections  from \cite{1996ApJ...465..487V}.  This model provided statistically a very poor fit for all the observations, and data residuals clearly indicated significant contribution from the accretion disk except for observation 1. 
Addition of a multi-color disk blackbody component \citep[\texttt{diskbb};][]{1984PASJ...36..741M} provided a marginal improvement in the fit statistics for the observations 2 to 5, whereas no improvement was noticed for the observation 1. We, therefore, have not considered the \texttt{diskbb} component for the observation 1 in our further analysis. Absence of disk contribution was also reported by \cite{2022MNRAS.510.4040H} for the same observation
during the 2019 outburst. However, with this model \texttt{constant*tbabs(diskbb+powerlaw)}, we noticed significant residuals at $\sim6-8$ keV and $>12$ keV region, indicating the presence of iron emission line and reflection hump in the spectra from all the observations. 

\begin{figure}
    \includegraphics[width=\columnwidth]{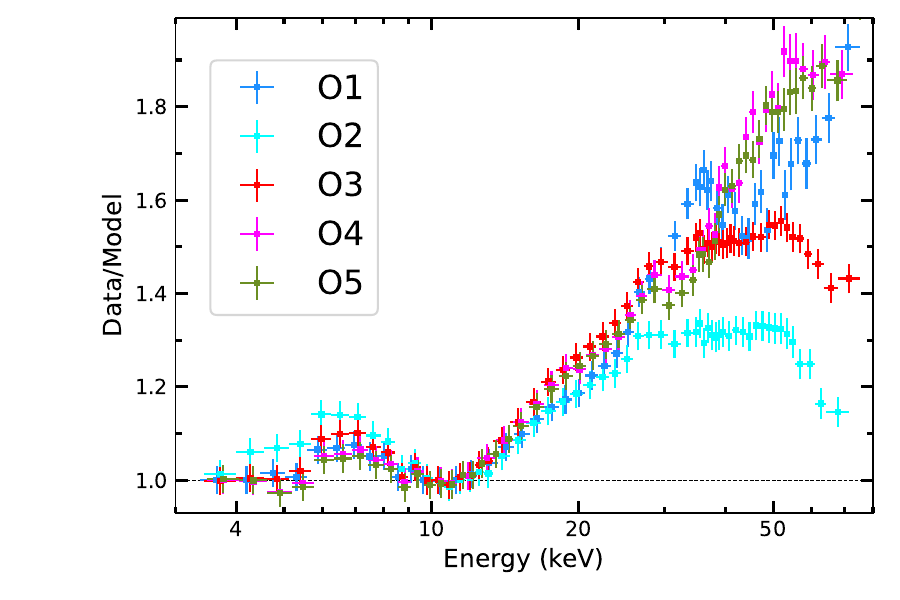}
    \caption{Ratio of LAXPC20 spectral data in $3-80$ keV to the fit with \texttt{tbabs(powerlaw+diskbb)}. Spectral fitting is performed 
    only in the $3-5$ and $9-12$ keV bands. Presence of the iron line excess and reﬂection hump are clearly visible in all the data set.}
    \label{fig:lxp_res}
\end{figure}

\begin{figure*}
    \centering
    \includegraphics[width=0.3\linewidth]{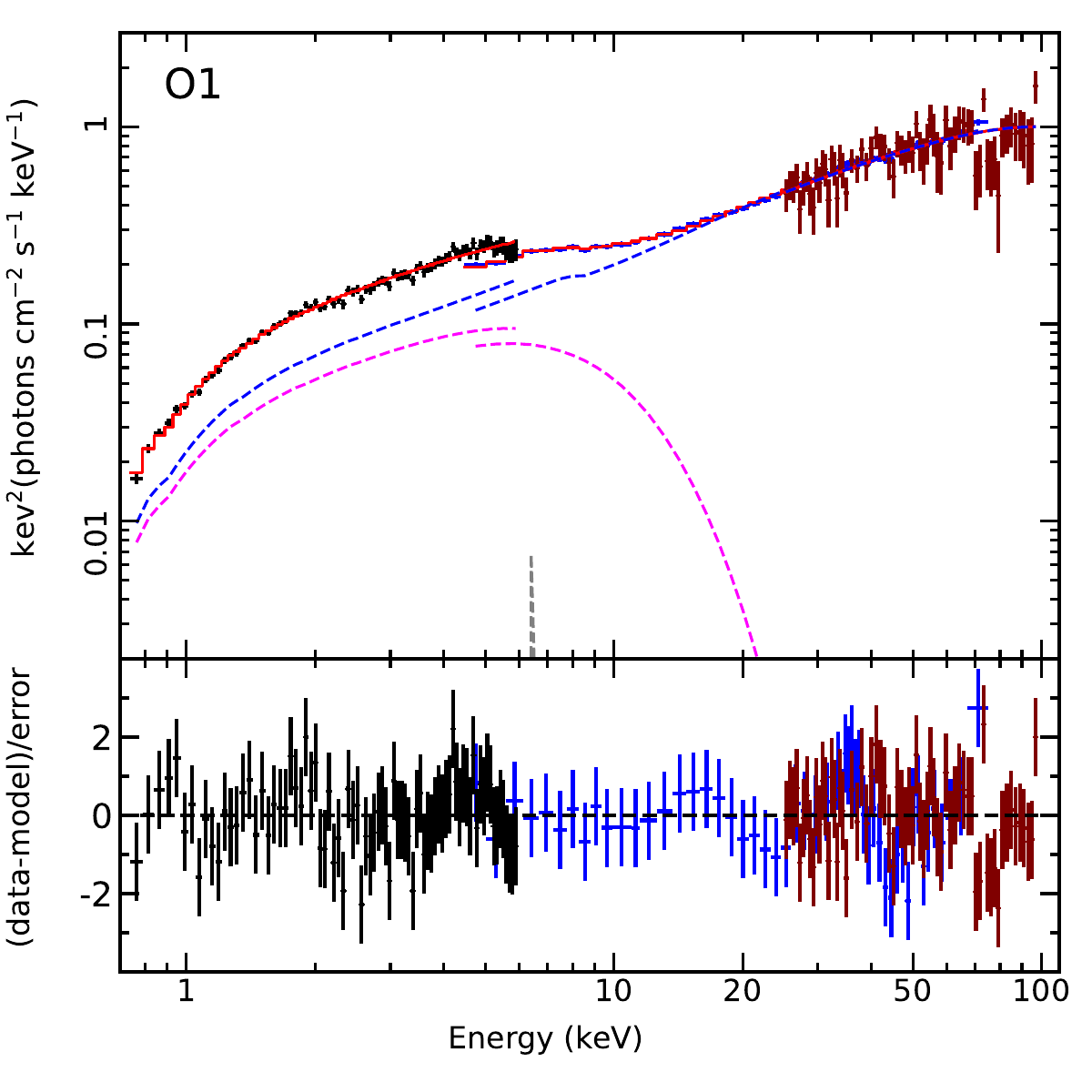}
    \includegraphics[width=0.3\linewidth]{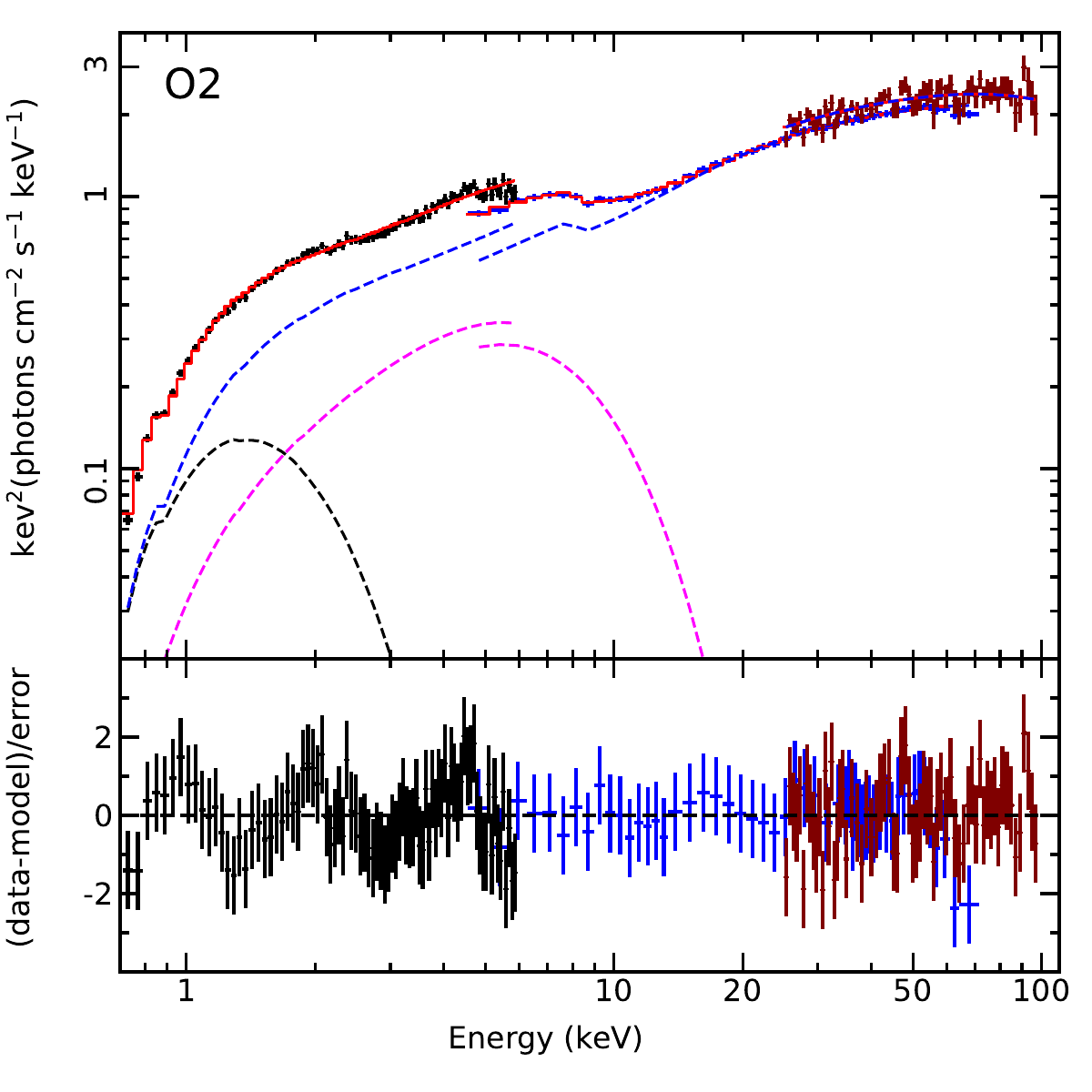}
    \includegraphics[width=0.3\linewidth]{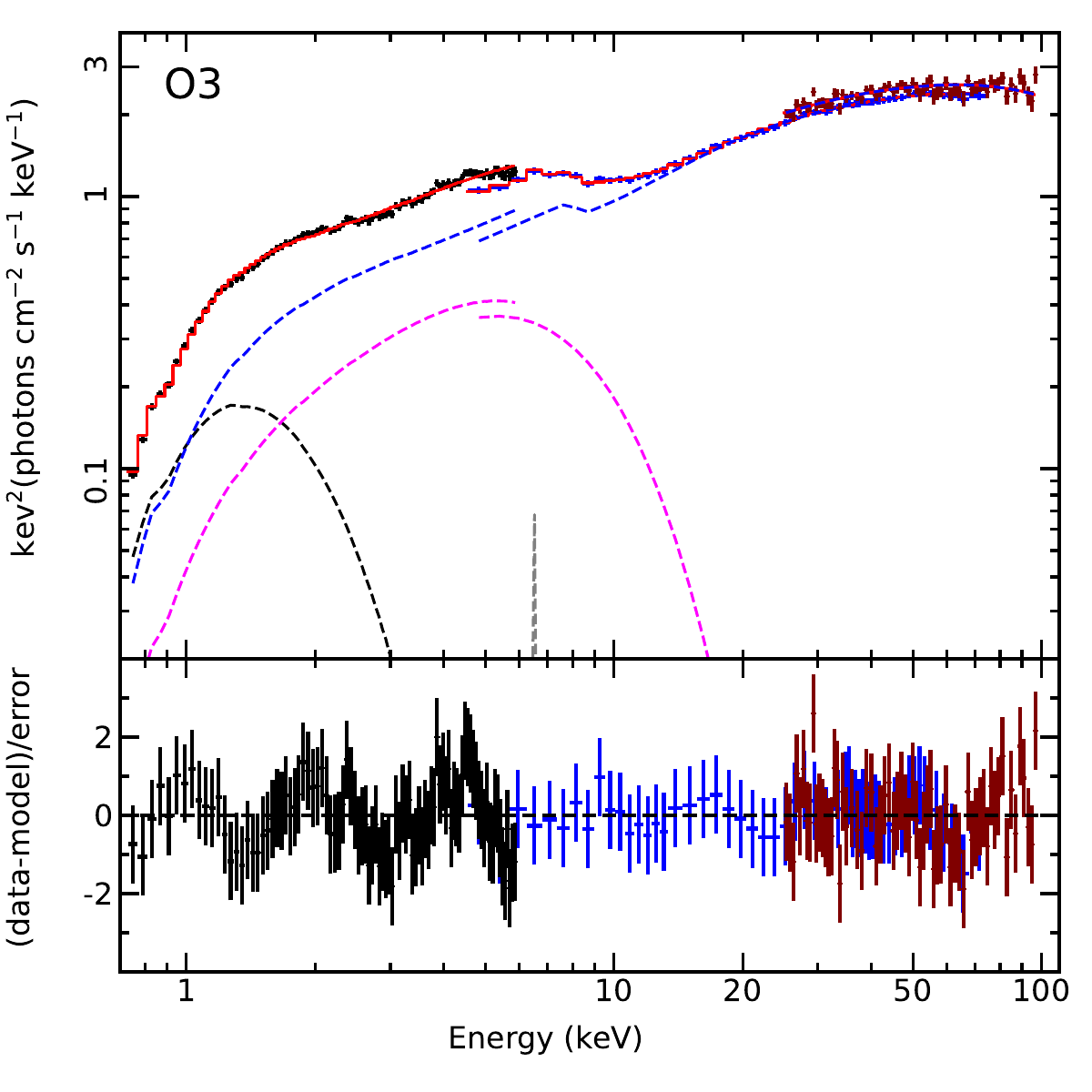}
    \includegraphics[width=0.3\linewidth]{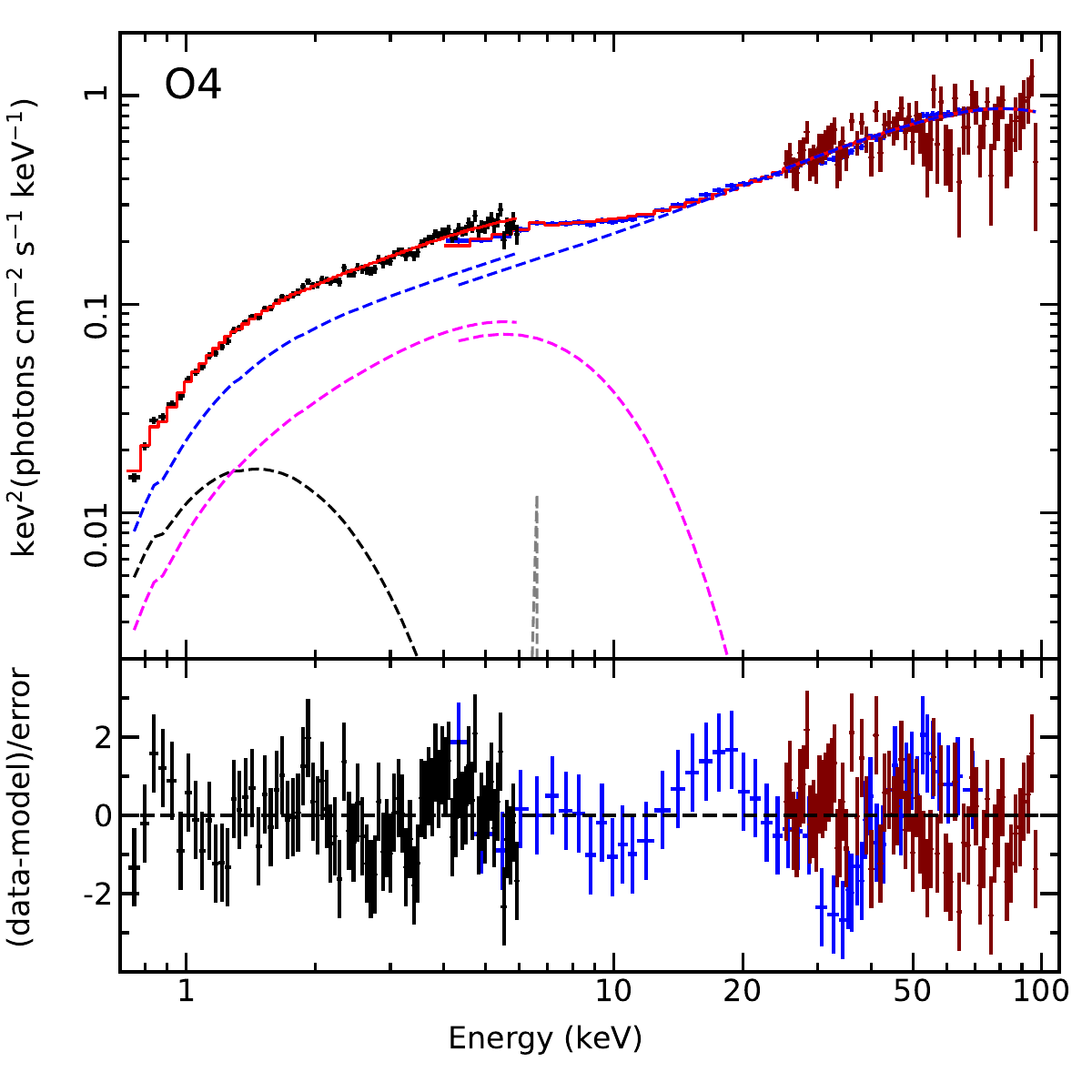}
    \includegraphics[width=0.3\linewidth]{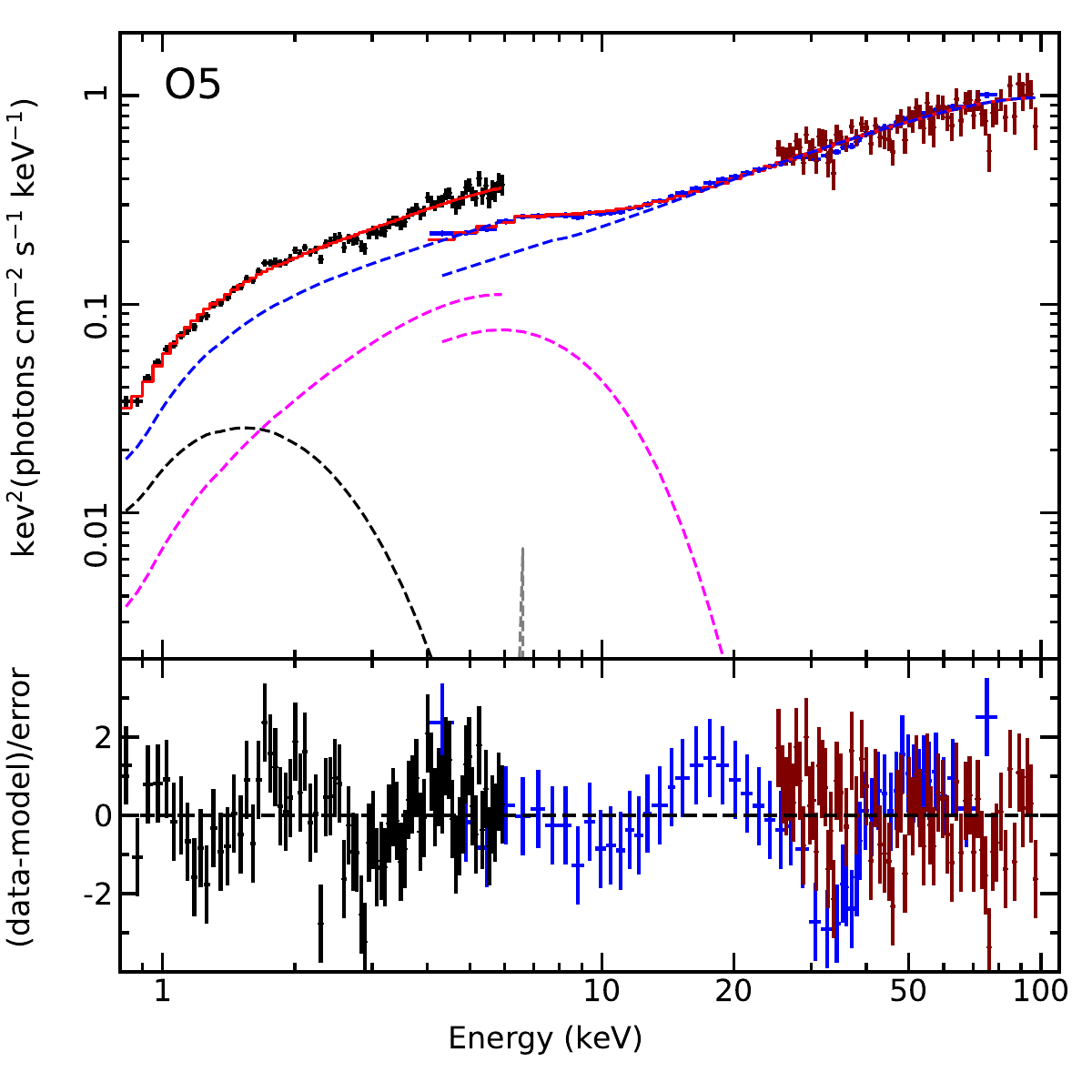}
  
    \caption{Joint SXT, LAXPC20 and CZTI spectral data ﬁtted with Model-1 for all the observations. Black, blue and maroon points 
    indicate the SXT, LAXPC20 and CZTI spectral data, respectively. The solid red lines indicate the extrapolated models of the 
    continuum ﬁt. The black and grey dashed lines indicate the \texttt{diskbb} and \texttt{gaussian} components, respectively. The \texttt{nthcomp1} and \texttt{nthcomp2} components are shown by blue and magenta colored dashed lines, respectively.}
    \label{fig:Model-1}
\end{figure*}

To show the presence of the iron line and reflection hump clearly, we first fitted the LAXPC20 spectral data in the $3-5$ keV and $9-12$ keV bands with the above mentioned model, and then compared the observed data with the continuum model. Figure~\ref{fig:lxp_res} shows the ratios of the LAXPC20 spectral data and the model for all the observations, where the excess due to iron  line and reflection hump are evident. We note here that the observations used in this work were mostly
taken at the rising phase of the outbursts (see Figure~\ref{fig:maxi_ltcrv}), where there is fair chance of the source to be in the hard state 
(described below). Since the accretion disk in BHTs is not extend down to the ISCO during
quiescence period, it is expected to be truncated away from the ISCO at the early phases of outbursts \citep{1996A&A...314..813L, 2001A&A...373..251D, 2018A&A...614A..79P}. In this scenario, the iron line should appear as a narrow emission feature, originating at a large distance from the BH. Keeping this in mind, we incorporated a \texttt{gaussian} component for the iron line, where the line energy and width were kept fixed at $6.5$ keV and $0.05$ keV. Another reason for keeping the line energy and width to be fixed is the low spectral resolution of the LAXPC20 detector. Besides, we also replaced the \texttt{powerlaw} component with the \texttt{nthcomp} that describes the shape of the continuum originating due to the thermal Comptonization of the soft disk photons from the accretion disk \citep{1996MNRAS.283..193Z}. We tied the seed photon temperature of the \texttt{nthcomp} component with the inner disk temperature ($kT_\mathrm{in}$) of the \texttt{diskbb} component, and kept it fixed at $0.2$ keV for observation 1. In addition, to account for the reflection from ionized material, we convolved \texttt{ireflect}\citep{1995MNRAS.273..837M} with the \texttt{nthcomp}, where \texttt{ireflect} is the generalized convolution form of \texttt{pexriv}, and describes only the hard X-ray shape of the reflected spectrum excluding all the line emissions. Since \texttt{ireflect} is a convolution component, we extended the sample energy range up to $1000$ keV. As the parameters of this component, we kept the reflection scaling factor (RS) free to vary and fixed all other parameters such as iron and other element abundances at their Solar values, inclination at $60^\circ$, disk temperature at $30000$ K and disk ionization parameter at a high value of 800. Irrespective of the addition of the \texttt{ireflect} component, significant residuals from all the observations were still visible, which could arise if some portion of the soft disk seed photons are getting Comptonized in a separate Comptonization region other than the first one. Hence, we incorporated an additional \texttt{nthcomp} component, independent of the first one, into our model. Similar to the first, the temperature of input seed photons of the second Comptonization region was tied with $kT_\mathrm{in}$, whereas it was fixed at $0.2$ keV in the case of observation 1. The photon index, electron temperature and the normalization were allowed to vary independently from the first Comptonization region. Addition of this second \texttt{nthcomp} component provided a substantial improvement in the fit statistics (at least $\Delta\chi^2=-101$ for two degrees of freedom). Aside from the aforementioned,  Xenon K emission and absorption like features at $\sim31-33$ keV can be noticed in the residuals of the LAXP20 spectral data from some of the observations. However, these features are not significant and are absent in the CZTI spectral data. Presence of such features in the LAXPC20 spectral data are mostly due to instrument calibration issues \citep{2019MNRAS.487.4221S}, and removing them had no effect on our best-fit spectral parameters. Thus our model \texttt{constant*tbabs(diskbb+ireflect*nthcomp+gaussian \hspace{0.5cm}+nthcomp)} (hereafter Model-1) is now adequate to describe 
the full X-ray energy band in a considerably manner with acceptable fit statistics. We note that small residuals in some of the observations are due to the calibration uncertainties of the respective instruments, and are not significant as they are mostly within $\pm~2\sigma$ level. We did not observe any significant affect of these residuals on the best-fit spectral parameters. Moreover, the value of the reduced $\chi^2$ shows an increase for O4 and O5, where the Xenon absorption feature becomes more prominent in comparison to the other observations. Removing this feature did not show any significant change in the best-fit spectral parameters. The best-fit spectral parameters obtained from Model-1 are listed in Table~\ref{tab:Model-1}, whereas the best-fit spectral models along with data are shown in Figure~\ref{fig:Model-1}. 

The Galactic absorption column density, $N_{\rm{H}}$, obtained from this Model-1, is found to vary between $\sim0.5-0.6\times10^{22}$ cm$^{-2}$ for all the observations, which is in good agreement with the findings of the previous studies. The inner disc temperature ($kT_\mathrm{in}$) is nearly the same within errors and ranges between  $\sim0.4$ and $0.5$ keV. However, the disk normalization for observations 4 and 5 is found to be significantly low in comparison to that for the observations 2 and 3. The photon index ($\Gamma_1$) and electron temperature ($kT_\mathrm{{e1}}$) obtained for the first Comptonization component appear to be consistent with what is typically expected in the LHS of BHTs. For the second Comptonization region, the photon index ($\Gamma_2$) is slightly lower than that of the first region for observations 2 to 5, whereas it is
slightly higher for observation 1. As the electron temperature ($kT_\mathrm{{e2}}$) was not constrained except for observation 1, we kept it fixed at the best-fit value for 
the rest of the observations. The best-fit values of $kT_\mathrm{{e2}}$ ($\lesssim2$ keV) indicate that the second Comptonization region is relatively much  colder than the first one. Furthermore, with the exception of observation  1, the normalization of the \texttt{nthcomp2} is found to be slightly lower than that of \texttt{nthcomp1}. The reflection scaling factor (RS) appears to be nearly similar within errors for all the observations, whereas it is fixed for observation 4
at the best-fit value due to difficulty in constraining. To compare the physical nature of these two different Comptonizing regions, we calculated the
optical depths ($\tau$) from the best-fit photon index ($\Gamma$) and electron temperatures ($kT_{\rm{e}}$) using the following relation
\citep{1996MNRAS.283..193Z, 1999MNRAS.309..561Z}:
\begin{equation}
    \tau = \sqrt{\frac{9}{4} + \frac{3m_e c^2}{kT_e [(\Gamma + \frac{1}{2})^2 - \frac{9}{4}]}} -\frac{3}{2}
\end{equation}
 where $m_e$ denotes the mass of the electron and $c$ to be the speed of the light. The derived values of the optical depth ($\tau$) for all the observations are listed in Table~\ref{tab:Model-1}. We find that $\tau$ for the second Comptonization region is very high with respect to that of the first one, implying the second
 region to be highly optically thick. 
 Apart from this, we also computed the Compton y-parameter ($y$) using the relation \citep{1996MNRAS.283..193Z}:
 \begin{equation}
     y = \frac{4kT_e}{m_e c^2} max(\tau, \tau^2)
 \end{equation}
 and found it to be higher for the second region than the first one. The unabsorbed flux estimated in the $0.7-100$ keV band is found to be higher by
 an order of magnitude for the observations 2 and 3 than that for the rest of the  three observations (see Table~\ref{tab:Model-1}). This suggests that the source was in a brighter hard state during the observations 2 and 3, taken in the course of 2021 outburst. We have also noticed that contribution from the disk is larger for observations 2 and 3 than
 for observations 4 and 5.  

\begin{figure*}
    \centering
    \includegraphics[width=0.3\linewidth]{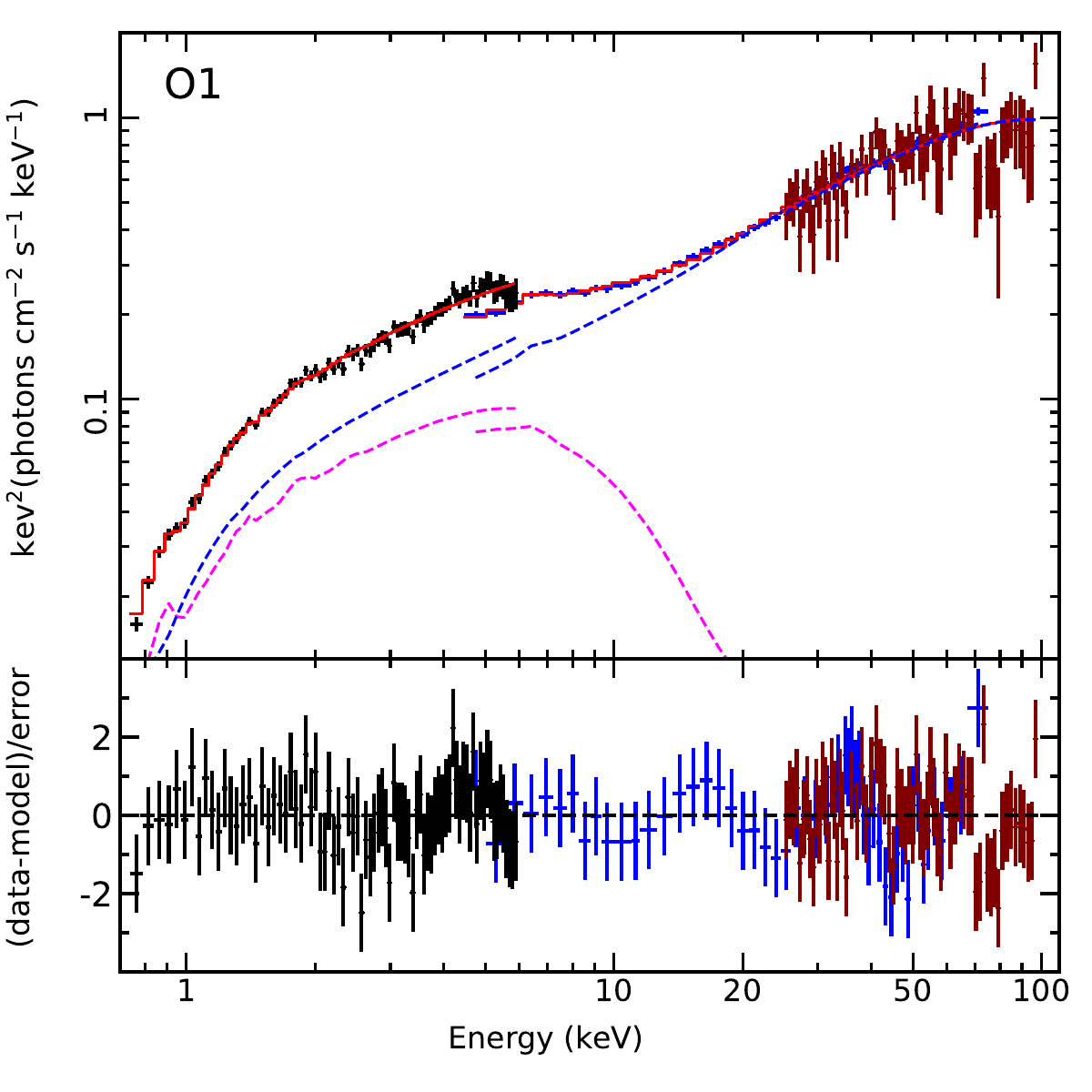} 
    \includegraphics[width=0.3\linewidth]{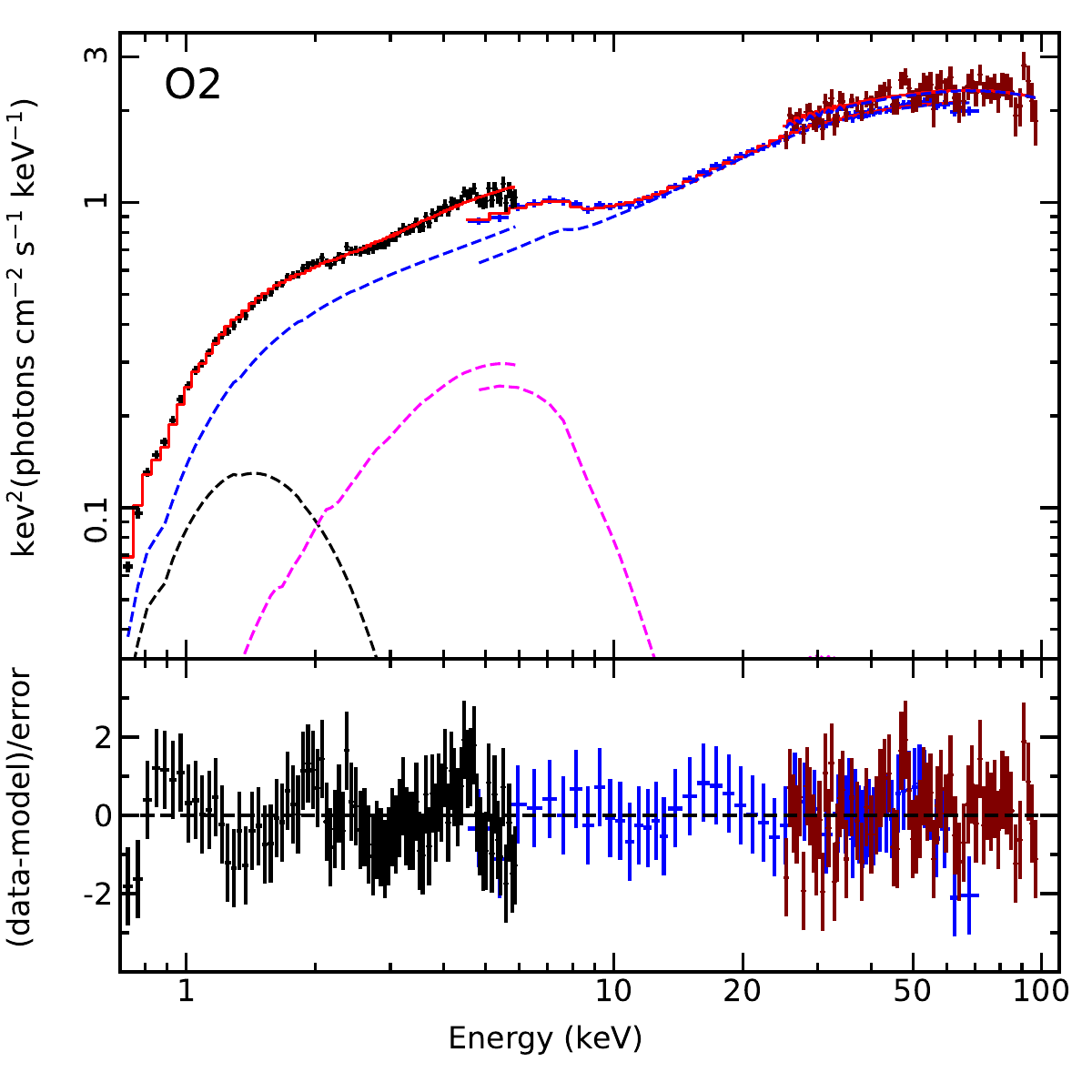}
    \includegraphics[width=0.3\linewidth]{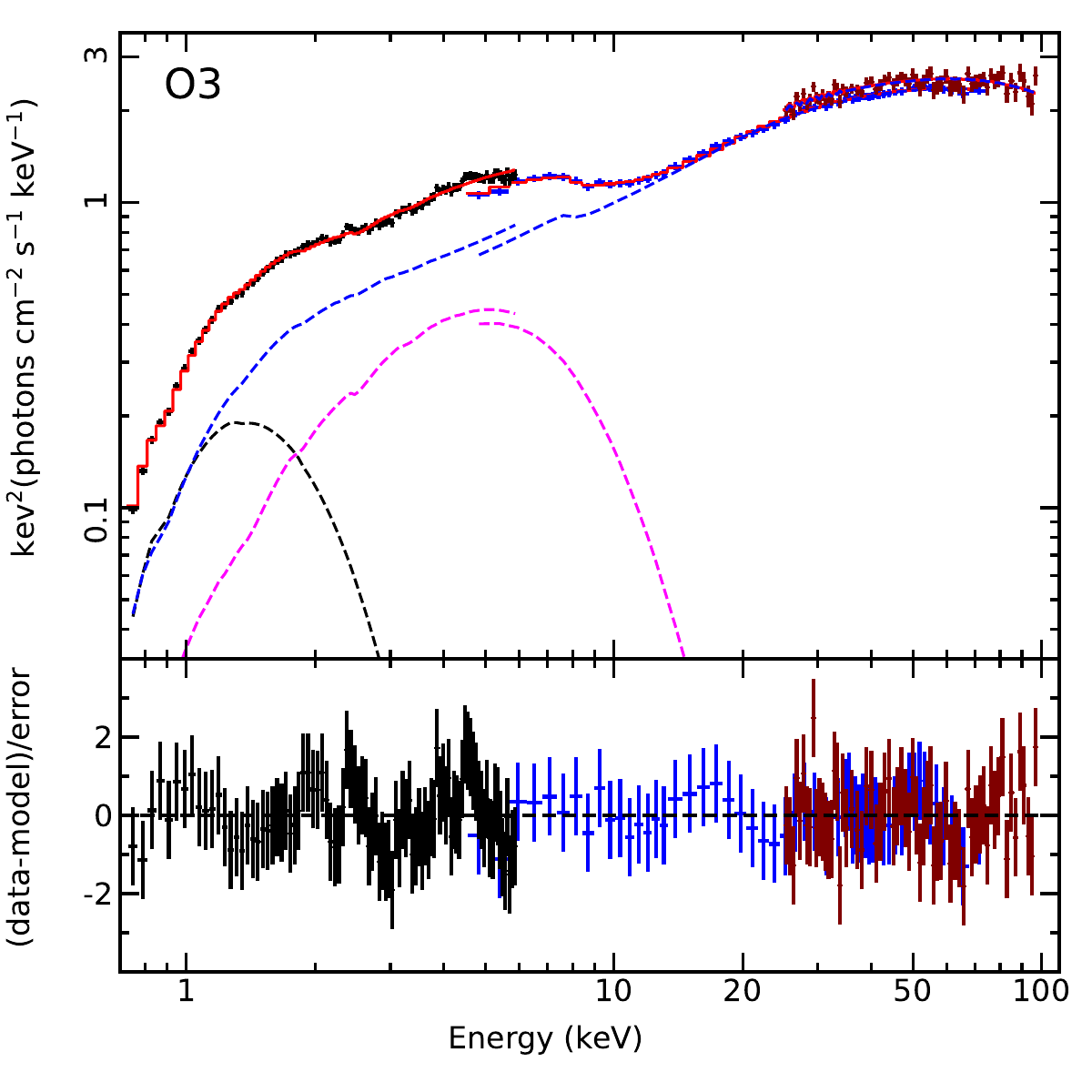}
    \includegraphics[width=0.3\linewidth]{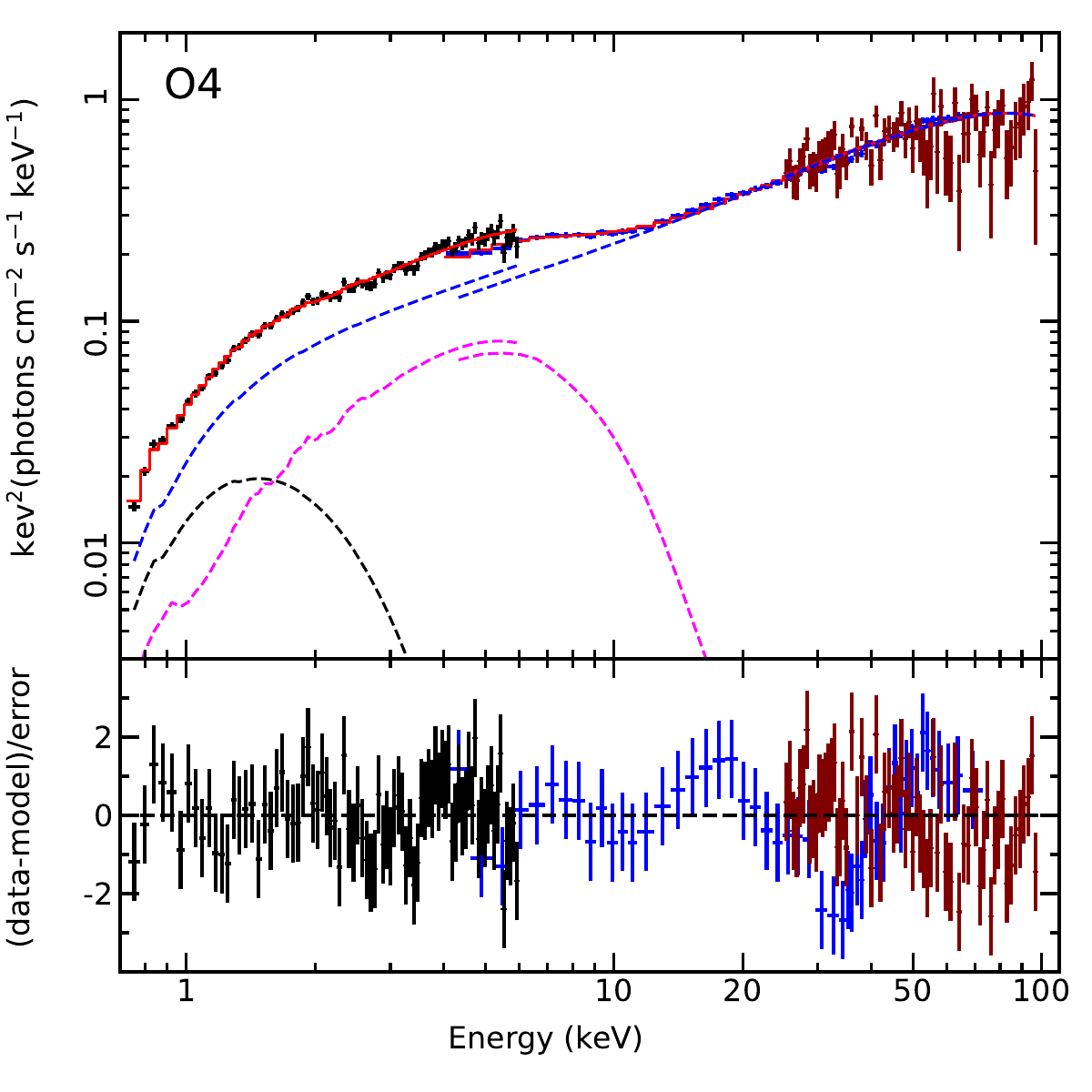}
    \includegraphics[width=0.3\linewidth]{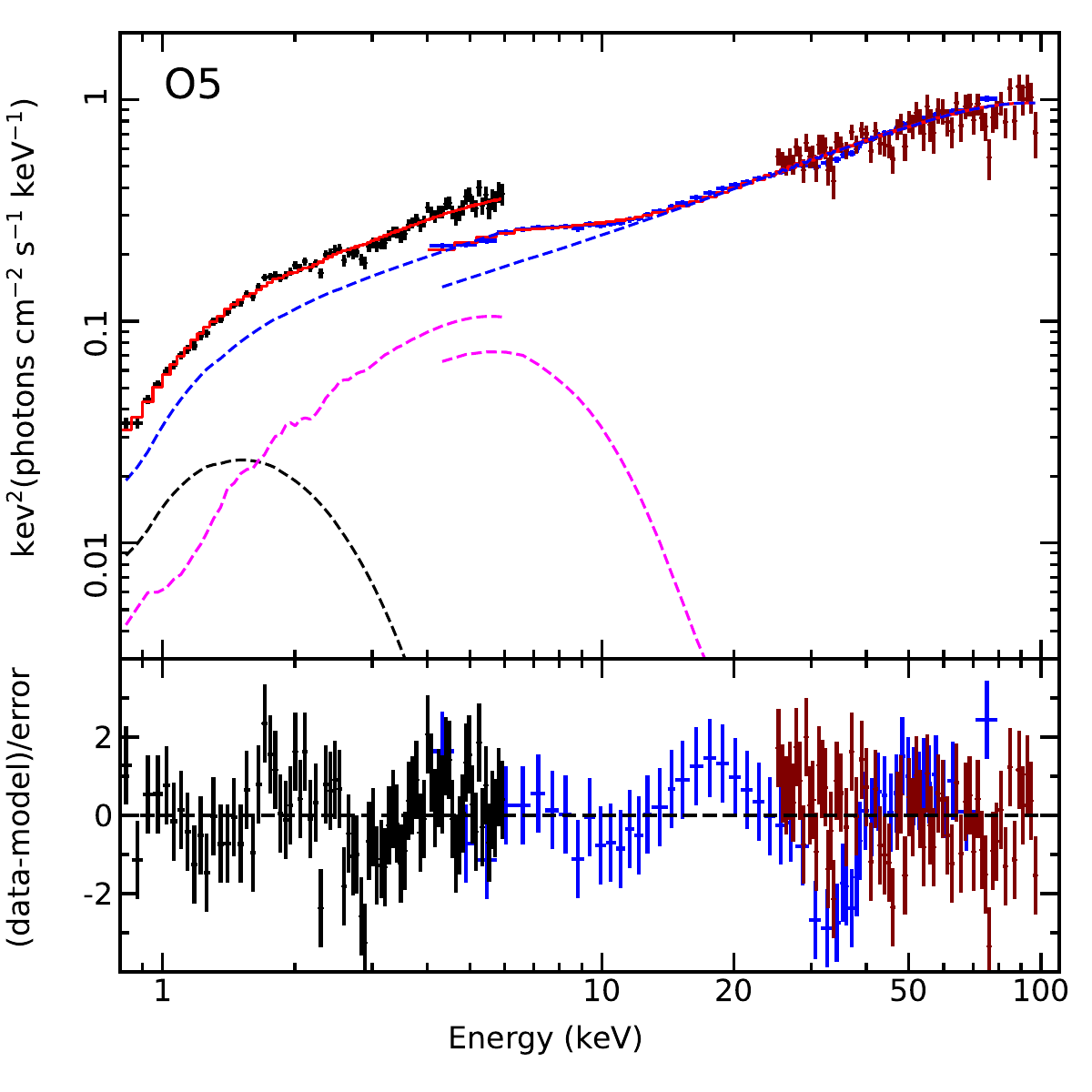}
  
    \caption{Joint SXT, LAXPC20 and CZTI spectral data ﬁtted with Model-2 for all the observations. Black, blue and maroon points 
    indicate the SXT, LAXPC20 and CZTI spectral data, respectively. The solid red lines indicate the extrapolated models of the 
    continuum ﬁt. The black dashed line indicates the \texttt{diskbb} component. The \texttt{relxillCp1} and \texttt{relxillCp2} 
    components are shown by blue and magenta colored dashed lines, respectively.}
    \label{fig:Model-2}
\end{figure*}

Further, we performed reflection spectroscopy in the framework of double Comptonization model similar to Model-1. For this purpose, we have used the relativistic reﬂection component \texttt{relxillCp}, which comes from the family of the blurred reﬂection model \texttt{relxill} \citep{2014ApJ...782...76G, 2014MNRAS.444L.100D}. The incident irradiation continuum in \texttt{relxillCp} is assumed to be a Comptonization spectrum that is identical to \texttt{nthcomp}. Initially we started fitting the data by replacing \texttt{ireflect}, \texttt{gaussian} and two \texttt{nthcomp} components with a single \texttt{relxillCp} component in our Model-1. The spin of the BH was fixed at $0.95^{+0.02}_{-0.08}$, derived by \cite{2016ApJ...821L...6P} using the very high state observations, which is equivalent to the ISCO radius of $\approx1.94~r_\mathrm{g}$. We also fixed disk inclination angle at $60^\circ$, iron abundance at the Solar value \citep{2016MNRAS.458.2199B} and outer disk radius at $\mathrm{1000 r_g}$. During the spectral fitting a single emissivity profile ($\propto r^{-q}$, $q$ being the emissivity index) was considered throughout the whole accretion disk, and $q$ was kept fixed at the canonical value of 3. Other parameters such as photon index ($\Gamma^\dag$), electron temperature ($kT^\dag_\mathrm{e}$), reflection fraction ($R_\mathrm{{ref}}$) and normalization were kept free and allowed to vary. However, we noticed that a single \texttt{relxillCp} 
component could not describe the spectra and significant residuals were observed in all the observations similar to what we found while using a single
\texttt{nthcomp} component. Thus, we included an additional \texttt{relxillCp} component in this model, and tied the parameters between these
two components except the photon index ($\Gamma^\dag$), electron temperatures ($kT^\dag_\mathrm{e}$), reflection fractions ($R_\mathrm{{ref}}$) and normalization. Inclusion of the additional \texttt{relxillCp} component provided a significant improvement in the fit-statistics 
(at least $\Delta \chi^2 = -100$ for four degrees of freedom). Here, we tied the inner disk radius ($R_\mathrm{{in}}$) between the first and second \texttt{relxillCp} components, and made the log of the ionization ($\mathrm{log\xi}$) parameter common for both the regions. We note that the contribution from the second Comptonization region is weaker than the first region (see Figure~\ref{fig:Model-1}) and hence, tying these parameters between these two regions will not affect the best-fit spectral parameters. Since the $R_\mathrm{{ref}}$ of the first Comptonization region for observations 4 and 5 was pegging to a very low value, we fixed it at $0.01$. We also kept the $R_\mathrm{{ref}}$ of the second region fixed at unity for observations 4 and 5. 
Thus our model \texttt{constant$\times$tbabs(diskbb+relxillCp+relxillCp)}, referred as Model-2, describes the spectra very well and provides better fits than the Model-1 in terms of $\chi^2/\mathrm{dof}$. It is to be noted here that we also performed the spectral  fitting by varying the inclination angle between $40^\circ-60^\circ$, and found it to be pegging at the higher value. Moreover, it required a significantly higher value of the iron abundance, much larger than the Solar. This may be due to an artifact of the current reflection models, where keeping the electron density of the disk fixed at $10^{15}~\mathrm{cm^{-3}}$ leads to an artificially high value of the iron abundance  \citep{2018ASPC..515..282G, 2022ApJ...928...11Z}. This issue has been addressed by \cite{2019MNRAS.484.1972J}, where they found iron abundance close to the Solar considering high-density reflection models. However, in our work, we keep the reflector density fixed at the model default value ($10^{15}~\mathrm{cm^{-3}}$), and hence the inclination and iron abundance at $60^\circ$ and Solar, respectively. Similar to Model-1, we also observe here small residuals in some observations due to calibration uncertainties. Additionally, the increase in reduce $\chi^2$ for O4 and O5 is similar to Model-1 (see above). The best-fit spectral parameters obtained from Model-2 are listed in Table~\ref{tab:Model-2}, whereas the best-fit models along with spectral data are shown in Figure~\ref{fig:Model-2}. The nature of the accretion disk in terms of inner disk temperature ($kT^\dag_\mathrm{in}$) and disk normalization is similar to what we have found with Model-1. The best-fit values of $\Gamma^\dag_1$ and $kT^\dag_\mathrm{e1}$ obtained for the first Comptonization region also suggest the source was in the hard state during all the observations. Using the Model-2, we could now constrain the electron temperature ($kT^\dag_{\rm{{e2}}}$) for the second Comptonization region, which is substantially cooler than the first region and consistent with the values obtained from the Model-1. Although the inner disk radius could not be constrained properly, the best-fit values indicate that the disk is truncated by several $r_\mathrm{g}$ for all the observations. The nature of the photon index ($\Gamma^\dag_2$) and electron temperature ($kT^\dag_\mathrm{e2}$) for the second Comptonization region from Model-2 is in agreement with those obtained from Model-1. Apart from this, the normalization parameter of the first Comptonization region appears to be higher by an order of magnitude than for the second region. 

\begin{table*}
	\caption{Best-fit broadband X-ray spectral parameters derived using Model--1}
    \centering 
   \begin{tabular}{lcccccc}
       \hline
      \hline
	Component & Parameter & O1 & O2 & O3 & O4 & O5 \\
	\hline
	TBabs & $N_{\rm{H}}$(10$^{22}$ cm$^{-2})$ & $0.47^{+0.04}_{-0.03}$ & 0.56$^{+0.07}_{-0.05}$ & 0.61$\pm0.05$ & 0.49$^{+0.08}_{-0.05}$ & 0.48$^{+0.05}_{-0.06}$ \\ \\
	
		Diskbb & $kT_\mathrm{in}$ (keV) & \nodata & 0.38$^{+0.05}_{-0.06}$ & 0.35$\pm0.04$  & 0.43$^{+0.09}_{-0.11}$ & 0.47$^{+0.12}_{-0.08}$ \\
		          & norm & \nodata & 1764.7$^{+2652.5}_{-768}$ & $3625.3^{+3251.8}_{-1497.1}$ & 114.6$^{+131}_{-53}$ & 116.8$^{+144.5}_{-63.0}$ \\ \\

	ireflect & RS & $0.33^{+0.62}_{-0.27}$ & 0.66$^{+0.19}_{-0.21}$ & 0.68$^{+0.20}_{-0.17}$ & 0.02$^f$ & $0.12^{+0.14}_{}$\\ \\
	
	Gaussian & norm ($\times 10^{-3}$) & $0.1^{+0.6}_{}$ & $<2.5$ & 1$^f$ & $0.2^{+0.5}_{}$ & $0.12^{+0.74}_{}$ \\ \\
			
	 NthComp1 & $\Gamma_1$ & $1.41\pm0.02$ & 1.62$\pm0.02$ & 1.63$\pm0.02$ & 1.43$\pm0.01$ & 1.45$\pm0.01$\\
	            & $kT_{\rm{e1}}$ & $43.3^{+479.6}_{-13.}$ & 65.35$^{+127.6}_{-22.9}$ & 56.09$^{+39.66}_{-13.91}$ & $30.3^{+4.7}_{-3.2}$ & $45.85^{+44.24}_{-10.73}$ \\
	            & norm & $0.05\pm0.01$ & 0.29$^{+0.04}_{-0.02}$ & 0.34$\pm0.03$ & 0.052$^{+0.006}_{-0.005}$ & 0.074$^{+0.008}_{-0.009}$ \\ \\
             
        NthComp2 & $\Gamma_2$ & $1.59^{+0.06}_{-0.08}$ & 1.41$^{+0.12}_{-0.11}$ & 1.46$\pm0.08$ & 1.39$\pm0.1$ & 1.27$^{+0.13}_{-0.16}$ \\
	            & $kT_{\rm{e2}}$ & $1.94^{+0.49}_{-0.33}$ & 1.62$^f$ & 1.62$^f$ & 1.62$^f$ & 1.62$^f$ \\
	            & norm & $0.04^{+0.02}_{-0.01}$ & 0.08$^{+0.04}_{-0.03}$ & 0.12$^{+0.04}_{-0.03}$ & 0.019$^{+0.009}_{-0.007}$ & 
             0.016$\pm0.009$ \\ \\
             \hline
        Cross-calibration & $C_{\rm{laxpc20}}$ & $0.85\pm0.04$ & $0.87^{+0.03}_{-0.04}$ & $0.92\pm0.03$ & $0.89\pm0.03$ & $0.69\pm0.03$ \\ 
                          & $C_{\rm{czti}}$ & $0.83^{+0.06}_{-0.05}$ & $0.94\pm0.04$ & $0.98\pm0.03$ & $0.89\pm0.05$ & $0.69\pm0.03$ \\ 
             
        \hline
            & $\chi^2$/dof & $166/188$ & $140.4/195$ & $131.2/206$ & $217.5/189$ & $246.5/187$\\
        \hline             
            & & & Derived Values & & \\
            \hline
    $F_\mathrm{{total}}$ [$10^{-9} \mathrm{erg~cm}^{-2} \mathrm{s}^{-1}$] &  & 3.4  & 11.2 & 12.6 & 3.1 & 4.2 \\
    $F_\mathrm{{disk}}$ [$10^{-11} \mathrm{erg~cm}^{-2} \mathrm{s}^{-1}$] &  & \nodata & 38.1 & 56.3 & 4.6 & 7.1 \\
    Disk Fraction ($\%$) & & \nodata & 3.5 & 4.5 & 1.5 & 1.7 \\
    $L/L_\mathrm{{Edd}}$ & & 0.02 & 0.07 & 0.08 & 0.02 & 0.03 \\ \\
    
    $F_\mathrm{{bol}}$ (total)  & & $4.02$ & $14.2$ & $15.8$ & $4.32$ & $6.24$ \\
    $F_\mathrm{{bol}}$ (nthcomp1) & & $4.0 (99\%)$ & $12.57 (88\%)$ & $13.38 (85\%)$ & $3.98 (92\%)$ & $5.86 (94\%)$ \\
    $L_{\mathrm{bol}}/L_{\mathrm{Edd}}$ & & 0.024 & 0.086 & 0.096 & 0.026 & 0.038 \\ \\

    & $\tau_1$ & $\sim3.7$ & $\sim2.1$ & $\sim2.27$ & $\sim4.5$ & $\sim3.38$\\
    & $\tau_2$ & $\sim17.87$ & $\sim24.55$ & $\sim22.9$ & $\sim25.3$ & $\sim31.3$ \\
    \\

    & $y$-par1 & $\sim4.64$ & $\sim2.26$ & $\sim2.26$ & $\sim4.8$ & $\sim4.1$ \\
    & $y$-par2 & $\sim4.84$ & $\sim7.6$ & $\sim6.65$ & $\sim8.1$ & $\sim12.4$ \\
             \\
	            
    \hline 
    \end{tabular}
    \label{tab:Model-1}
    \begin{tablenotes}
    \item Notes-- $F_\mathrm{{total}}$, $F_\mathrm{{disk}}$ and $L$ are total unabsorbed flux, disk flux and luminosity in the $0.7-100$ keV band, respectively. $F_\mathrm{{bol}}$ represents bolometric flux and $L_{\mathrm{bol}}$ the bolometric luminosity in the
    $0.001-500$ keV band, and are in the unit of $\mathrm{10^{-9}~erg~cm^{-2}~s^{-1}}$. $\tau_1$ and $\tau_2$ are the optical depths, and $y$-par1
     and $y$-par2 are the Compton y-parameter of the two different Comptonization regions. $f$ indicates the fixed parameters.
    \end{tablenotes}
\end{table*}

\begin{table*}
	\caption{Best-fit broadband X-ray spectral parameters derived using Model--2}
   \centering
   \begin{tabular}{l@{\hskip 0.2in}c@{\hskip 0.2in}c@{\hskip 0.2in}c@{\hskip 0.2in}c@{\hskip 0.2in}cc}
 
      \hline
      \hline
	Component & Parameter & O1 & O2 & O3 & O4 & O5 \\
	\hline
	TBabs & $N_{\rm{H}}$(10$^{22}$ cm$^{-2})$ & $0.55^{+0.06}_{-0.05}$ & 0.63$^{+0.11}_{-0.06}$ & 0.69$\pm0.03$ & $0.56^{+0.08}_{-0.05}$ & $0.56^{+0.06}_{-0.05}$\\ \\

		DISKBB & $kT_{\rm{in}}$ (keV) & \nodata & 0.38$^{+0.05}_{-0.08}$ & 0.35$^{+0.04}_{-0.05}$ & $0.42^{+0.17}_{-0.12}$ &  $0.44^{+0.10}_{-0.08}$ \\
		& norm & \nodata & 1788.4$^{+4263.5}_{-816.6}$ & 4607.6$^{+5283.5}_{-1819.9}$ & $164.6^{+331.8}_{-106.2}$ & $155.3^{+256.4}_{-98.7}$ \\ \\

	RelxillCp1 
                & $\Gamma^\dag_1$ & $1.40\pm0.03$ & 1.64$\pm0.05$ & 1.60$^{+0.05}_{-0.03}$ & $1.44^{+0.01}_{-0.02}$ & $1.45\pm0.01$ \\
                & $kT^\dag_{\rm{{e1}}}$ (keV) & $37.1^{+36.1}_{-8.3}$ & $>37.7$ & $31.5^{+5.1}_{-3.8}$ & $50.0^{+10.6}_{-8.5}$ & 
      $39.1^{+7.3}_{-4.9}$ \\
		    & $R_{\rm{in}}$ ($r_{\mathrm{g}}$) & $>42.6$ & $<11.29$ & $<9.4$ & $>44.6$ & $>13.2$ \\
		    
		    & $log \xi$ & $1.3^{+0.4}_{-0.2}$ & $<1.86$ & $<3.05$ & $1.7^{+0.96}_{-0.23}$  & $1.85^{+0.55}_{-0.36}$ \\
		      & $R_{\mathrm{ref1}}$ & $0.18^{+0.34}_{-0.15}$ & $0.55^{+0.36}_{-0.28}$  & $0.68^{+0.27}_{-0.23}$ & $0.01^f$ & $0.01^f$ \\
		    & norm1 ($\times10^{-3}$)& $5.24^{+0.7}_{-0.5}$ & 14.7$^{+1.6}_{-0.9}$ & 13.7$^{+1.1}_{-0.6}$ & $4.7^{+0.4}_{-0.3}$ & $6.9^{+0.5}_{-0.4}$ \\ \\

        RelxillCp2 & $\Gamma^\dag_2$ & $1.66^{+0.07}_{-0.09}$ & $<1.62$ & $<1.3$ & $<1.47$ & $<1.34$ \\
		    & $kT^\dag_{\rm{{e2}}}$ (keV) & $2.01^{+0.56}_{-0.33}$ & $1.33^{+0.49}_{-0.11}$ &  1.45$^{+0.24}_{-0.12}$ & $1.5^{+0.2}_{-0.1}$ & 1.49$^{+0.13}_{-0.11}$ \\
		    & $R_{\mathrm{ref2}}$ & $2.0\pm1.3$ & $>0.8$ & $1.26^{+1.23}_{-0.63}$ & $1^f$ & $1^f$ \\
		    & norm2 ($\times10^{-4}$)& $4.8^{+2.0}_{-1.1}$ & 6$^{+6}_{-1}$ & 10.7$^{+3.0}_{-2.8}$ & $2.5^{+1.0}_{-0.4}$ & $3^{+0.7}_{-0.4}$ \\ \\

      \hline
        Cross-calibration & $C_{\rm{laxpc20}}$ & $0.85\pm0.04$ & $0.85\pm0.03$ & $0.90\pm0.03$ & $0.87\pm0.03$ & $0.68\pm0.03$ \\ 
                          & $C_{\rm{czti}}$ & $0.82\pm0.05$ & $0.93\pm0.04$ & $0.97\pm0.03$ & $0.88\pm0.05$ & $0.68\pm0.03$ \\
                          \hline
      
		    & $\chi^2$/dof & $160.2/186$ & $136.8/192$ & $118.3/202$ & $207/187$ & $235.8/186$ \\
	\hline 
    \end{tabular}
    \label{tab:Model-2}
    \begin{tablenotes}
    \item Notes-- $f$ indicates the fixed parameters.
    \end{tablenotes}
\end{table*}

\subsection{Power Density Spectra}

\begin{figure}
    \includegraphics[width=\columnwidth]{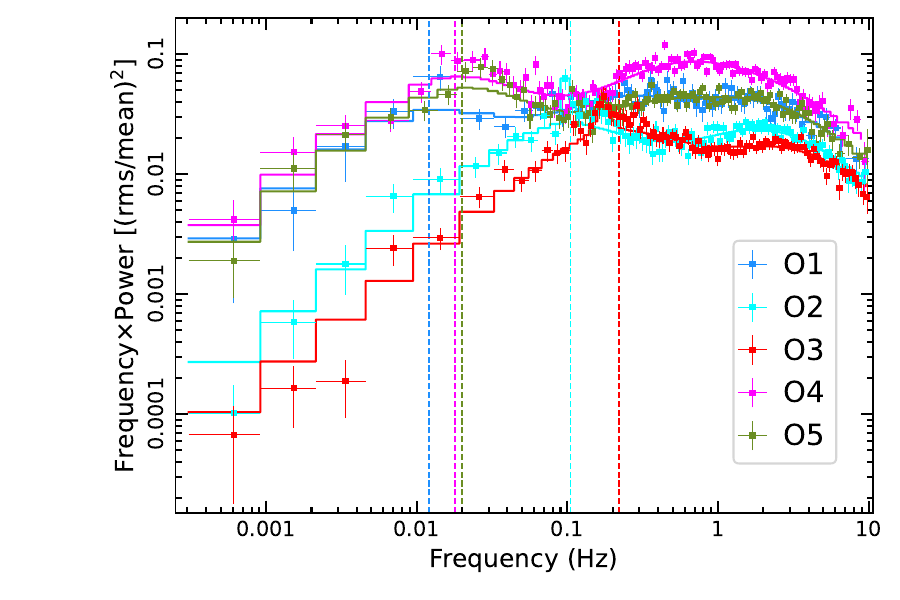}
    \caption{Power density spectra derived for all the five observations in the $3-15$ keV band. The solid lines indicate the best-fit total model to each PDS. The dashed vertical lines represent the break frequency for each observations.}
    \label{fig:brk_freq}
\end{figure}

\begin{figure}
    \includegraphics[width=\columnwidth]{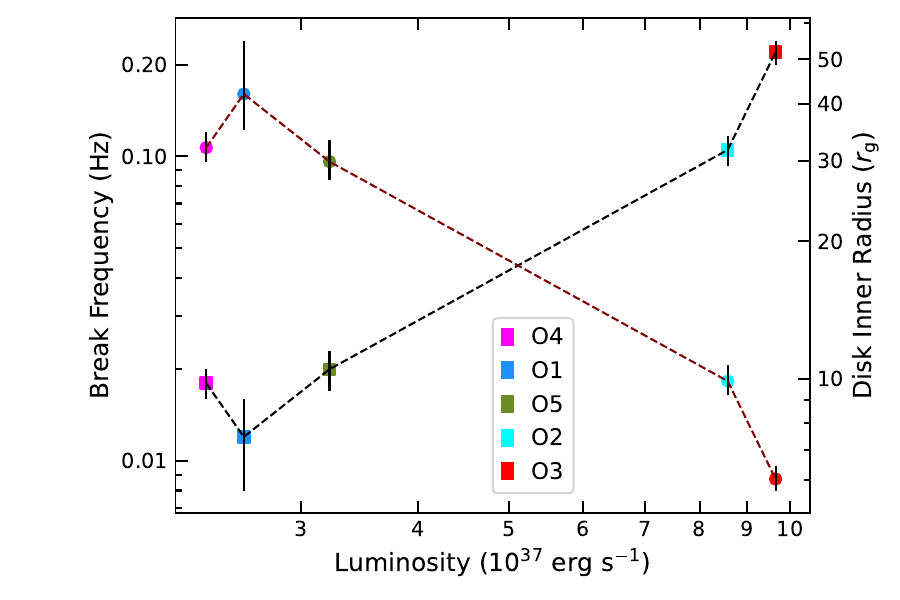}
    \caption{Evolution of the break frequency (squares) and the corresponding inner disk radius (circles) with the source luminosity in the $0.7-100$ keV band. Each data point represents a single observation.}
    \label{fig:brkFreq_lum}
\end{figure}

In the LHS, the PDSs can be described with a combination of several Lorentzian profiles required for peaked and broadband noise components, also termed as band-limited noise (BLN) components. The characteristics frequency of a Lorentzian component can be defined as $\nu_\mathrm{c}=\sqrt{\nu^2+\Delta^2}$, where $\nu$ and $\Delta$ are the centroid frequency and half-width at half-maximum (HWHM), respectively \citep{1997A&A...322..857B}. For the BLN components (where quality factor $Q<1$) 
$\nu=0$, and thus $\nu_\mathrm{c}$ becomes equal to $\Delta$, which is also called as the break frequency ($\nu_\mathrm{brk}$) \citep[see][for details]{2002ApJ...572..392B}. In this work, we derived rms-normalized PDSs from the LAXPC20 data from each observation in the 3-15 keV band using the \texttt{laxpc\_find\_freqlag}{\footnote{\url{http://astrosat-ssc.iucaa.in/laxpcData}}} task, which also accounts for the correction of LAXPC dead-time
 and the corresponding background count rates \citep{2015ApJ...800..109B, 2016ApJ...833...27Y, 2018MNRAS.477.5437A, 2022ApJ...933...69C}. To derive the PDSs, we used the total LAXPC20 exposure time for all the observations except for observations 3 and 5, where the PDSs were extracted from the orbits taken on the first day of the observations due to their very large exposure times. We found that the PDSs from the observations having 
 relatively low luminosity (observations O1, O4 and O5) can be described with three zero-centered Lorentzians required for the BLN components. On the other hand,
the PDSs from the observations O2 and O3, where the luminosity is comparatively higher than the other three observations, exhibit two BLN components along with the presence of a quasi-periodic oscillation (QPO) at $\sim0.1$ Hz and $\sim0.17$ Hz, respectively. We fitted the PDSs of these two observations with two zero-centered Lorentzians and a single peaked Lorentzian to account for the BLN components and QPO, respectively. We then derived the break frequency ($\nu_{\mathrm{brk}}$) from each of the best-fit PDS using the first zero-centered Lorentzian Component. Figure~\ref{fig:brk_freq} shows the PDSs and the best-fit models for all the observations, whereas the corresponding break frequencies of all the PDSs are listed in Table~\ref{tab:pds_table}. From the shape of the PDSs, it is clear that the variability power is higher for the observations with
 lower luminosity than those having higher luminosity. We also notice that the break frequency gradually increases with the increase in the 
 source luminosity (see Figure~\ref{fig:brkFreq_lum}).

\section{Discussion and Concluding Remarks} \label{sec:discusiion}

We have performed broadband spectral analysis of the five hard/hard-intermediate state observations of the low-mass BHT GX~339--4 taken during 
the $2019-2022$ outbursts by AstroSat, where the spectral data from all the three co-aligned X-ray instruments are utilized. Almost all of these observations were made while the source was in the rising phase of its corresponding outbursts (see Figure~\ref{fig:maxi_ltcrv}). However, the HIDs derived using Swift/XRT observations (see Figure~\ref{fig:xrt_hid}) suggest that the nature of these outbursts was not always the same. The outbursts in 2019 and 2022 appear to be failed or hard-only outbursts, whereas the 2021 outburst is found to be a successful one. The reason for an outburst not entering the soft state is still an open question and beyond the scope of this study. \cite{2021MNRAS.507.5507A} studied the observational differences between a failed and full outburst using a sample of low-mass BHXRBs in the X-ray and optical/infrared (IR) bands. Their findings suggest that a failed outburst can be predicted prior to its onset, when an increase in the source brightness in the optical/IR bands is observed due to a higher mass accretion rate. Another study by \cite{2023ApJ...958..153L} has found that the BHTs exhibit a systematic evolution in the X-ray variability a few weeks before the onset of canonical outbursts, which is not seen in the cases of failed outbursts. They also proposed that the variability and state transitions in BHTs are caused by changes in the turbulence in the outer region of the disk. Figure~\ref{fig:xrt_hid} further depicts that the source was in the LHS during all the observations except for O3, where it might have made a transition to the HIMS,  as a significant decrease in the hardness ratio along with an increase in the count rates was observed.

In our spectral analysis, we find that the broadband spectra can be described with two separate Comptonization components having different electron
temperatures and optical depths, as well as the  accretion disk contributing at low energies. The first Comptonization component appears to be relatively harder with higher electron temperature and lower optical depth with respect to the second component, which is much softer with a low electron temperature and very high optical depth. Moreover, this hard component is found to dominate both the total unabsorbed flux and bolometric luminosity (see Table~\ref{tab:Model-1}), and also mainly responsible for the origin of the narrow reflection features in the X-ray spectra. The Photon index ($\Gamma_1$ and $\Gamma^\dag_1$) obtained for this region from both the models indicates that the source was in the LHS or HIMS (for O3) during all the observations. We notice that $\Gamma_1$ and $\Gamma^\dag_1$ are higher for O2 and O3, observed close to the peak of the 2021, when the source luminosity was higher than the other observations (see Figure~\ref{fig:maxi_ltcrv} and Table~\ref{tab:Model-1}). The spectral steepening with the increase in luminosity indicates that the source began its spectral transition over the course of O2 to O3, and then finally entered into the HIMS during O3 (see Figure~\ref{fig:xrt_hid}). The estimated inner radius of the accretion disk from the reflection spectroscopy using Model-2 (see Table~\ref{tab:Model-2}) indicates that the disk is significantly truncated away from the ISCO for all the observations, when the luminosity varies in the range of $\sim2-8\%$ $L_\mathrm{{Edd}}$ during the LHS (and the HIMS for O3). The inner accretion disk deviating from the ISCO in the bright LHS has also been reported by \cite{2021ApJ...909L...9Z}, where they found substantially truncated disk at the source luminosity of $\sim5-15\%L_\mathrm{{Edd}}$, considering a double corona geometry for the BHT MAXI~J1820+070. However, the study by \cite{2019ApJ...885...48G} during the 2017 failed outburst of GX~339--4 suggested that the inner disk is required to be in proximity to the ISCO, particularly in the initial stages, when the source luminosity is around $\gtrsim 1\%L_\mathrm{E}$. Thus, it is evident that the extent of the inner accretion disk at the LHS of BHTs may not strictly depend upon the source luminosity. Furthermore, we notice that the disk truncation decreases as the source brightness or luminosity increases (see Table~\ref{tab:Model-1} and \ref{tab:Model-2}). Similar kind of characteristics of the disk inner radius with the source luminosity has been reported by \cite{2015A&A...573A.120P} and \cite{2016MNRAS.458.2199B}. We find the accretion disk to be low to moderately ionized, which is expected in the truncated disk scenario. The normalization parameters of the Comptonization regions from both Model-1 and 2 suggest that the second region is comparatively weaker than the first region for almost all the observations.

\begin{table}[]
    \caption{Derived break frequencies from the best-fit PDSs of each observation.}
    \centering
    \begin{tabular}{lccc}
    \hline
    Obs No & Date of Obs & $\nu_\mathrm{brk}$ (Hz) & $R_\mathrm{{in}}$ ($r_{\mathrm{_g}}$) \\
    \hline
        O1 & 2019-09-22 & $0.012\pm0.004$ & $42^{+13}_{-7}$\\
        O2 & 2021-02-13 & $0.105\pm0.012$ & $9.89^{+0.83}_{-0.69}$ \\
        O3 & 2021-03-02 & $0.22\pm0.02$ & $6.04^{+0.40}_{-0.34}$ \\
        O4 & 2022-09-07 & $0.018\pm0.002$ & $32.05^{+2.62}_{-2.17}$ \\
        O5 & 2022-09-09 & $0.020\pm0.003$ & $29.88^{+3.41}_{-2.66}$ \\
      \hline
    \end{tabular}
    \label{tab:pds_table}
\end{table}

The inner disk temperature remains low throughout all the observations and varies between $\sim0.4-0.5$ keV. The disk normalization goes down to  lower values in the case of the low-luminosity observations with respect to the observations at higher luminosity. We note here that our assumption of the soft excess as
disk blackbody may not be realistic. Similar soft excess component in GX~339--4 was also reported by \cite{1998MNRAS.301..435Z} and \cite{2002MNRAS.337..829W}, where 
they suggested the origin of the soft excess to be due to non-homogeneity in the geometry of the corona and accretion disk. Besides, the best-fit values of $N_{\rm{H}}$
from both the models are also well consistent with the previously reported values \citep{2015A&A...573A.120P, 2015ApJ...813...84G, 2019ApJ...885...48G, 2023MNRAS.521.3570Y}, which also indicates that the soft excess is not an artefact due to the Galactic absorption column density. However, we also note here that
this soft excess has an effect in making the spectrum from the second Comptonization region artificially hard despite of its low electron temperature for the observations 2 to 5. On the other hand, the spectrum from the second comptonization region, where the contribution from the soft excess is negligible, appears to be 
much steeper than the first Comptonization region.

As can be seen from Fig.~\ref{fig:brkFreq_lum}, the break frequency increases with the increase in the source luminosity. In addition, 
\cite{1999A&A...352..182G} observed that the break frequency ($\mathrm{\nu_{brk}}$) is strongly correlated with the steepening of the power-law component and increasing amplitude of the reflection component, which can be associated with the decreasing  disk truncation radius as a result of surge in the penetration 
of the cool accretion disk into the hot inner flow \citep{2001MNRAS.321..759C}. Also, the $\mathrm{\nu_{brk}}$ is related to the viscous time scale 
($\mathrm{t_{vis}}$),  given as \citep{2002apa..book.....F, 2007A&ARv..15....1D}
\begin{equation}
    t_{\mathrm{vis}}=4.5~ \alpha^{-1}~ (H/R)^{-2}~ (R/6R_g)^{3/2}~ (M_{\rm{BH}}/10M_\odot)~ \mathrm{ms}
\end{equation}
where $\alpha$ is the dimensionless viscosity parameter \citep{1973A&A....24..337S}, $H$ is the vertical scale height of the disk, and $M_{\rm{BH}}$ is the mass
 of the BH. Using the above equation and the values of the $\nu_\mathrm{brk}$ mentioned in Table~\ref{tab:pds_table}, we estimated the disk truncation radius for each observation considering $\alpha=0.1$ and $H/R=0.1$ for thin disk, and $M_{\rm{BH}}=10M_\odot$. 
 Table~\ref{tab:pds_table} lists the estimated inner disk radius from the $\nu_\mathrm{brk}$ for all the five observations, where it can be seen that the truncation radius decreases with the increase of the $\mathrm{\nu_{brk}}$. However, the inner accretion disk radius remains truncated for all observations up to a several $r_\mathrm{g}$. It is to be noted here that these estimates of the inner disk radius are independent of the 
 disk inclination angle, and are also in well agreement with those obtained through the reflection spectroscopy using Model-2 (see Table~\ref{tab:Model-2}). In addition, Figure~\ref{fig:brkFreq_lum} also depicts that the inner disk radius, estimated from $\nu_{\mathrm{brk}}$, decreases with the increase in the source luminosity, which also validates our findings of $R_\mathrm{{in}}$ from the reflection spectroscopy using Model-2.


\section{Acknowledgments}
We thank the anonymous referee for the useful comments that have improved the quality of the paper. This publication uses data from the AstroSat mission of the Indian Space Research Organisation (ISRO), archived at the Indian Space Science Data Centre (ISSDC). This work has used data from the SXT, LAXPC and CZTI instruments on board AstroSat. The LAXPC data were processed by the Payload Operation Center (POC) at TIFR, Mumbai. This work has been performed utilizing the calibration databases and auxiliary analysis tools developed, maintained, and distributed by the AstroSat-SXT team with members from various institutions in India and abroad and the SXT POC at the TIFR, Mumbai (\url{https://www.tifr.res.in/~astrosat_sxt/index.html}). The SXT data were processed and veriﬁed by the SXT POC.  CZT-Imager is built by a consortium of institutes across India, including the TIFR, Mumbai; the Vikram Sarabhai Space Centre, Thiruvananthapuram; ISRO Satellite Centre (ISAC), Bengaluru; Inter-University Centre for Astronomy and Astrophysics (IUCAA), Pune; Physical Research Laboratory, Ahmedabad; and Space Application Centre, Ahmedabad. Contributions from the vast technical team from all these institutes are gratefully acknowledged.  This research has also made use of data supplied by the UK Swift Science Data Centre at the University of Leicester and MAXI data provided by RIKEN, JAXA, and the MAXI team. AAZ acknowledges support from the Polish National Science Center under the grants 2019/35/B/ST9/03944 and 2023/48/Q/ST9/00138, and from the Copernicus Academy under the grant CBMK/01/24.

%

\vspace{5mm}
\facilities{AstroSat(SXT, LAXPC and CZTI), Swift(XRT), MAXI.}


\software{XSPEC\citep[12.13.0c;][]{1996ASPC..101...17A}, Julia, Matplotlib, LAXPCSoft, HEASoft(V. 6.31.1).}




\bibliography{reference}{}

\begin{thebibliography}{}
\expandafter\ifx\csname natexlab\endcsname\relax\def\natexlab#1{#1}\fi
\providecommand{\url}[1]{\href{#1}{#1}}
\providecommand{\dodoi}[1]{doi:~\href{http://doi.org/#1}{\nolinkurl{#1}}}
\providecommand{\doeprint}[1]{\href{http://ascl.net/#1}{\nolinkurl{http://ascl.net/#1}}}
\providecommand{\doarXiv}[1]{\href{https://arxiv.org/abs/#1}{\nolinkurl{https://arxiv.org/abs/#1}}}

\bibitem[{{Agrawal} {et~al.}(2017){Agrawal}, {Yadav}, {Antia}, {Dedhia},
  {Shah}, {Chauhan}, {Manchanda}, {Chitnis}, {Gujar}, {Katoch}, {Kurhade},
  {Madhwani}, {Manojkumar}, {Nikam}, {Pandya}, {Parmar}, {Pawar}, {Roy},
  {Paul}, {Pahari}, {Misra}, {Ravichandran}, {Anilkumar}, {Joseph},
  {Navalgund}, {Pandiyan}, {Sarma}, \& {Subbarao}}]{2017JApA...38...30A}
{Agrawal}, P.~C., {Yadav}, J.~S., {Antia}, H.~M., {et~al.} 2017, Journal of
  Astrophysics and Astronomy, 38, 30, \dodoi{10.1007/s12036-017-9451-z}

\bibitem[{{Agrawal} {et~al.}(2018){Agrawal}, {Nandi}, {Girish}, \&
  {Ramadevi}}]{2018MNRAS.477.5437A}
{Agrawal}, V.~K., {Nandi}, A., {Girish}, V., \& {Ramadevi}, M.~C. 2018, \mnras,
  477, 5437, \dodoi{10.1093/mnras/sty1005}

\bibitem[{{Alabarta} {et~al.}(2021){Alabarta}, {Altamirano}, {M{\'e}ndez},
  {C{\'u}neo}, {Vincentelli}, {Castro-Segura}, {Garc{\'\i}a}, {Luff}, \&
  {Veledina}}]{2021MNRAS.507.5507A}
{Alabarta}, K., {Altamirano}, D., {M{\'e}ndez}, M., {et~al.} 2021, \mnras, 507,
  5507, \dodoi{10.1093/mnras/stab2241}

\bibitem[{{Antia} {et~al.}(2017){Antia}, {Yadav}, {Agrawal}, {Verdhan Chauhan},
  {Manchanda}, {Chitnis}, {Paul}, {Dedhia}, {Shah}, {Gujar}, {Katoch},
  {Kurhade}, {Madhwani}, {Manojkumar}, {Nikam}, {Pandya}, {Parmar}, {Pawar},
  {Pahari}, {Misra}, {Navalgund}, {Pandiyan}, {Sharma}, \&
  {Subbarao}}]{2017ApJS..231...10A}
{Antia}, H.~M., {Yadav}, J.~S., {Agrawal}, P.~C., {et~al.} 2017, \apjs, 231,
  10, \dodoi{10.3847/1538-4365/aa7a0e}

\bibitem[{{Arnaud}(1996)}]{1996ASPC..101...17A}
{Arnaud}, K.~A. 1996, in Astronomical Society of the Pacific Conference Series,
  Vol. 101, Astronomical Data Analysis Software and Systems V, ed. G.~H.
  {Jacoby} \& J.~{Barnes}, 17

\bibitem[{{Bachetti} {et~al.}(2015){Bachetti}, {Harrison}, {Cook}, {Tomsick},
  {Schmid}, {Grefenstette}, {Barret}, {Boggs}, {Christensen}, {Craig},
  {Fabian}, {F{\"u}rst}, {Gandhi}, {Hailey}, {Kara}, {Maccarone}, {Miller},
  {Pottschmidt}, {Stern}, {Uttley}, {Walton}, {Wilms}, \&
  {Zhang}}]{2015ApJ...800..109B}
{Bachetti}, M., {Harrison}, F.~A., {Cook}, R., {et~al.} 2015, \apj, 800, 109,
  \dodoi{10.1088/0004-637X/800/2/109}

\bibitem[{{Banerjee} {et~al.}(2024){Banerjee}, {Dewangan}, {Knigge},
  {Georganti}, {Gandhi}, {Mithun}, {Saikia}, {Bhattacharya}, {Russell},
  {Lewis}, \& {Zdziarski}}]{2024arXiv240208237B}
{Banerjee}, S., {Dewangan}, G.~C., {Knigge}, C., {et~al.} 2024, arXiv e-prints,
  arXiv:2402.08237, \dodoi{10.48550/arXiv.2402.08237}

\bibitem[{{Basak} \& {Zdziarski}(2016)}]{2016MNRAS.458.2199B}
{Basak}, R., \& {Zdziarski}, A.~A. 2016, \mnras, 458, 2199,
  \dodoi{10.1093/mnras/stw420}

\bibitem[{{Belloni} {et~al.}(2005){Belloni}, {Homan}, {Casella}, {van der
  Klis}, {Nespoli}, {Lewin}, {Miller}, \& {M{\'e}ndez}}]{2005A&A...440..207B}
{Belloni}, T., {Homan}, J., {Casella}, P., {et~al.} 2005, \aap, 440, 207,
  \dodoi{10.1051/0004-6361:20042457}

\bibitem[{{Belloni} {et~al.}(2002){Belloni}, {Psaltis}, \& {van der
  Klis}}]{2002ApJ...572..392B}
{Belloni}, T., {Psaltis}, D., \& {van der Klis}, M. 2002, \apj, 572, 392,
  \dodoi{10.1086/340290}

\bibitem[{{Belloni} {et~al.}(1997){Belloni}, {van der Klis}, {Lewin}, {van
  Paradijs}, {Dotani}, {Mitsuda}, \& {Miyamoto}}]{1997A&A...322..857B}
{Belloni}, T., {van der Klis}, M., {Lewin}, W.~H.~G., {et~al.} 1997, \aap, 322,
  857

\bibitem[{{Casella} {et~al.}(2005){Casella}, {Belloni}, \&
  {Stella}}]{2005ApJ...629..403C}
{Casella}, P., {Belloni}, T., \& {Stella}, L. 2005, \apj, 629, 403,
  \dodoi{10.1086/431174}

\bibitem[{{Chand} {et~al.}(2020){Chand}, {Agrawal}, {Dewangan}, {Tripathi}, \&
  {Thakur}}]{2020ApJ...893..142C}
{Chand}, S., {Agrawal}, V.~K., {Dewangan}, G.~C., {Tripathi}, P., \& {Thakur},
  P. 2020, \apj, 893, 142, \dodoi{10.3847/1538-4357/ab829a}

\bibitem[{{Chand} {et~al.}(2022){Chand}, {Dewangan}, {Thakur}, {Tripathi}, \&
  {Agrawal}}]{2022ApJ...933...69C}
{Chand}, S., {Dewangan}, G.~C., {Thakur}, P., {Tripathi}, P., \& {Agrawal},
  V.~K. 2022, \apj, 933, 69, \dodoi{10.3847/1538-4357/ac7154}

\bibitem[{{Chattopadhyay} {et~al.}(2024){Chattopadhyay}, {Kumar}, {Rao},
  {Bhargava}, {Vadawale}, {Ratheesh}, {Dewangan}, {Bhattacharya}, {Mithun}, \&
  {Bhalerao}}]{2024ApJ...960L...2C}
{Chattopadhyay}, T., {Kumar}, A., {Rao}, A.~R., {et~al.} 2024, \apjl, 960, L2,
  \dodoi{10.3847/2041-8213/ad118d}

\bibitem[{{Churazov} {et~al.}(2001){Churazov}, {Gilfanov}, \&
  {Revnivtsev}}]{2001MNRAS.321..759C}
{Churazov}, E., {Gilfanov}, M., \& {Revnivtsev}, M. 2001, \mnras, 321, 759,
  \dodoi{10.1046/j.1365-8711.2001.04056.x}

\bibitem[{{Connors} {et~al.}(2022){Connors}, {Garc{\'\i}a}, {Tomsick},
  {Mastroserio}, {Grinberg}, {Steiner}, {Jiang}, {Fabian}, {Parker},
  {Harrison}, {Hare}, {Mallick}, \& {Lazar}}]{2022ApJ...935..118C}
{Connors}, R. M.~T., {Garc{\'\i}a}, J.~A., {Tomsick}, J., {et~al.} 2022, \apj,
  935, 118, \dodoi{10.3847/1538-4357/ac7ff2}

\bibitem[{{Dauser} {et~al.}(2014){Dauser}, {Garcia}, {Parker}, {Fabian}, \&
  {Wilms}}]{2014MNRAS.444L.100D}
{Dauser}, T., {Garcia}, J., {Parker}, M.~L., {Fabian}, A.~C., \& {Wilms}, J.
  2014, \mnras, 444, L100, \dodoi{10.1093/mnrasl/slu125}

\bibitem[{{Deegan} {et~al.}(2009){Deegan}, {Combet}, \&
  {Wynn}}]{2009MNRAS.400.1337D}
{Deegan}, P., {Combet}, C., \& {Wynn}, G.~A. 2009, \mnras, 400, 1337,
  \dodoi{10.1111/j.1365-2966.2009.15573.x}

\bibitem[{{Done} \& {Diaz Trigo}(2010)}]{2010MNRAS.407.2287D}
{Done}, C., \& {Diaz Trigo}, M. 2010, \mnras, 407, 2287,
  \dodoi{10.1111/j.1365-2966.2010.17092.x}

\bibitem[{{Done} \& {Gierli{\'n}ski}(2006)}]{2006MNRAS.367..659D}
{Done}, C., \& {Gierli{\'n}ski}, M. 2006, \mnras, 367, 659,
  \dodoi{10.1111/j.1365-2966.2005.09968.x}

\bibitem[{{Done} {et~al.}(2007){Done}, {Gierli{\'n}ski}, \&
  {Kubota}}]{2007A&ARv..15....1D}
{Done}, C., {Gierli{\'n}ski}, M., \& {Kubota}, A. 2007, \aapr, 15, 1,
  \dodoi{10.1007/s00159-007-0006-1}

\bibitem[{{Dubus} {et~al.}(2001){Dubus}, {Hameury}, \&
  {Lasota}}]{2001A&A...373..251D}
{Dubus}, G., {Hameury}, J.~M., \& {Lasota}, J.~P. 2001, \aap, 373, 251,
  \dodoi{10.1051/0004-6361:20010632}

\bibitem[{{Dunn} {et~al.}(2011){Dunn}, {Fender}, {K{\"o}rding}, {Belloni}, \&
  {Merloni}}]{2011MNRAS.411..337D}
{Dunn}, R.~J.~H., {Fender}, R.~P., {K{\"o}rding}, E.~G., {Belloni}, T., \&
  {Merloni}, A. 2011, \mnras, 411, 337,
  \dodoi{10.1111/j.1365-2966.2010.17687.x}

\bibitem[{{Esin} {et~al.}(1997){Esin}, {McClintock}, \&
  {Narayan}}]{1997ApJ...489..865E}
{Esin}, A.~A., {McClintock}, J.~E., \& {Narayan}, R. 1997, \apj, 489, 865,
  \dodoi{10.1086/304829}

\bibitem[{{Fabian} {et~al.}(1989){Fabian}, {Rees}, {Stella}, \&
  {White}}]{1989MNRAS.238..729F}
{Fabian}, A.~C., {Rees}, M.~J., {Stella}, L., \& {White}, N.~E. 1989, \mnras,
  238, 729, \dodoi{10.1093/mnras/238.3.729}

\bibitem[{{Frank} {et~al.}(2002){Frank}, {King}, \&
  {Raine}}]{2002apa..book.....F}
{Frank}, J., {King}, A., \& {Raine}, D.~J. 2002, {Accretion Power in
  Astrophysics: Third Edition}

\bibitem[{{Garc{\'\i}a} {et~al.}(2014){Garc{\'\i}a}, {Dauser}, {Lohfink},
  {Kallman}, {Steiner}, {McClintock}, {Brenneman}, {Wilms}, {Eikmann},
  {Reynolds}, \& {Tombesi}}]{2014ApJ...782...76G}
{Garc{\'\i}a}, J., {Dauser}, T., {Lohfink}, A., {et~al.} 2014, \apj, 782, 76,
  \dodoi{10.1088/0004-637X/782/2/76}

\bibitem[{{Garc{\'\i}a} {et~al.}(2018{\natexlab{a}}){Garc{\'\i}a}, {Kallman},
  {Bautista}, {Mendoza}, {Deprince}, {Palmeri}, \&
  {Quinet}}]{2018ASPC..515..282G}
{Garc{\'\i}a}, J.~A., {Kallman}, T.~R., {Bautista}, M., {et~al.}
  2018{\natexlab{a}}, in Astronomical Society of the Pacific Conference Series,
  Vol. 515, Workshop on Astrophysical Opacities, 282,
  \dodoi{10.48550/arXiv.1805.00581}

\bibitem[{{Garc{\'\i}a} {et~al.}(2015){Garc{\'\i}a}, {Steiner}, {McClintock},
  {Remillard}, {Grinberg}, \& {Dauser}}]{2015ApJ...813...84G}
{Garc{\'\i}a}, J.~A., {Steiner}, J.~F., {McClintock}, J.~E., {et~al.} 2015,
  \apj, 813, 84, \dodoi{10.1088/0004-637X/813/2/84}

\bibitem[{{Garc{\'\i}a} {et~al.}(2018{\natexlab{b}}){Garc{\'\i}a}, {Steiner},
  {Grinberg}, {Dauser}, {Connors}, {McClintock}, {Remillard}, {Wilms},
  {Harrison}, \& {Tomsick}}]{2018ApJ...864...25G}
{Garc{\'\i}a}, J.~A., {Steiner}, J.~F., {Grinberg}, V., {et~al.}
  2018{\natexlab{b}}, \apj, 864, 25, \dodoi{10.3847/1538-4357/aad231}

\bibitem[{{Garc{\'\i}a} {et~al.}(2019){Garc{\'\i}a}, {Tomsick}, {Sridhar},
  {Grinberg}, {Connors}, {Wang}, {Steiner}, {Dauser}, {Walton}, {Xu},
  {Harrison}, {Foster}, {Grefenstette}, {Madsen}, \&
  {Fabian}}]{2019ApJ...885...48G}
{Garc{\'\i}a}, J.~A., {Tomsick}, J.~A., {Sridhar}, N., {et~al.} 2019, \apj,
  885, 48, \dodoi{10.3847/1538-4357/ab384f}

\bibitem[{{Gilfanov} {et~al.}(1999){Gilfanov}, {Churazov}, \&
  {Revnivtsev}}]{1999A&A...352..182G}
{Gilfanov}, M., {Churazov}, E., \& {Revnivtsev}, M. 1999, \aap, 352, 182,
  \dodoi{10.48550/arXiv.astro-ph/9910084}

\bibitem[{{Heida} {et~al.}(2017){Heida}, {Jonker}, {Torres}, \&
  {Chiavassa}}]{2017ApJ...846..132H}
{Heida}, M., {Jonker}, P.~G., {Torres}, M.~A.~P., \& {Chiavassa}, A. 2017,
  \apj, 846, 132, \dodoi{10.3847/1538-4357/aa85df}

\bibitem[{{Homan} \& {Belloni}(2005)}]{2005Ap&SS.300..107H}
{Homan}, J., \& {Belloni}, T. 2005, \apss, 300, 107,
  \dodoi{10.1007/s10509-005-1197-4}

\bibitem[{{Husain} {et~al.}(2022){Husain}, {Misra}, \&
  {Sen}}]{2022MNRAS.510.4040H}
{Husain}, N., {Misra}, R., \& {Sen}, S. 2022, \mnras, 510, 4040,
  \dodoi{10.1093/mnras/stab3780}

\bibitem[{{Jiang} {et~al.}(2019){Jiang}, {Fabian}, {Wang}, {Walton},
  {Garc{\'\i}a}, {Parker}, {Steiner}, \& {Tomsick}}]{2019MNRAS.484.1972J}
{Jiang}, J., {Fabian}, A.~C., {Wang}, J., {et~al.} 2019, \mnras, 484, 1972,
  \dodoi{10.1093/mnras/stz095}

\bibitem[{{Kaastra} \& {Bleeker}(2016)}]{2016A&A...587A.151K}
{Kaastra}, J.~S., \& {Bleeker}, J.~A.~M. 2016, \aap, 587, A151,
  \dodoi{10.1051/0004-6361/201527395}

\bibitem[{{Kolehmainen} \& {Done}(2010)}]{2010MNRAS.406.2206K}
{Kolehmainen}, M., \& {Done}, C. 2010, \mnras, 406, 2206,
  \dodoi{10.1111/j.1365-2966.2010.16835.x}

\bibitem[{{Kolehmainen} {et~al.}(2014){Kolehmainen}, {Done}, \& {D{\'\i}az
  Trigo}}]{2014MNRAS.437..316K}
{Kolehmainen}, M., {Done}, C., \& {D{\'\i}az Trigo}, M. 2014, \mnras, 437, 316,
  \dodoi{10.1093/mnras/stt1886}

\bibitem[{{Lasota} {et~al.}(1996){Lasota}, {Narayan}, \&
  {Yi}}]{1996A&A...314..813L}
{Lasota}, J.~P., {Narayan}, R., \& {Yi}, I. 1996, \aap, 314, 813,
  \dodoi{10.48550/arXiv.astro-ph/9605011}

\bibitem[{{Lucchini} {et~al.}(2023){Lucchini}, {Ten Have}, {Wang}, {Homan},
  {Kara}, {Adegoke}, {Connors}, {Dauser}, {Garcia}, {Mastroserio}, {Ingram},
  {van der Klis}, {K{\"o}nig}, {Lewin}, {Mallick}, {Nathan}, {O'Neill},
  {Panagiotou}, {Piotrowska}, \& {Uttley}}]{2023ApJ...958..153L}
{Lucchini}, M., {Ten Have}, M., {Wang}, J., {et~al.} 2023, \apj, 958, 153,
  \dodoi{10.3847/1538-4357/ad0294}

\bibitem[{{Ludlam} {et~al.}(2015){Ludlam}, {Miller}, \&
  {Cackett}}]{2015ApJ...806..262L}
{Ludlam}, R.~M., {Miller}, J.~M., \& {Cackett}, E.~M. 2015, \apj, 806, 262,
  \dodoi{10.1088/0004-637X/806/2/262}

\bibitem[{{Magdziarz} \& {Zdziarski}(1995)}]{1995MNRAS.273..837M}
{Magdziarz}, P., \& {Zdziarski}, A.~A. 1995, \mnras, 273, 837,
  \dodoi{10.1093/mnras/273.3.837}

\bibitem[{{Markert} {et~al.}(1973){Markert}, {Canizares}, {Clark}, {Lewin},
  {Schnopper}, \& {Sprott}}]{1973ApJ...184L..67M}
{Markert}, T.~H., {Canizares}, C.~R., {Clark}, G.~W., {et~al.} 1973, \apjl,
  184, L67, \dodoi{10.1086/181290}

\bibitem[{{Miller} {et~al.}(2006){Miller}, {Homan}, {Steeghs}, {Rupen},
  {Hunstead}, {Wijnands}, {Charles}, \& {Fabian}}]{2006ApJ...653..525M}
{Miller}, J.~M., {Homan}, J., {Steeghs}, D., {et~al.} 2006, \apj, 653, 525,
  \dodoi{10.1086/508644}

\bibitem[{{Miller} {et~al.}(2008){Miller}, {Reynolds}, {Fabian}, {Cackett},
  {Miniutti}, {Raymond}, {Steeghs}, {Reis}, \& {Homan}}]{2008ApJ...679L.113M}
{Miller}, J.~M., {Reynolds}, C.~S., {Fabian}, A.~C., {et~al.} 2008, \apjl, 679,
  L113, \dodoi{10.1086/589446}

\bibitem[{{Mitsuda} {et~al.}(1984){Mitsuda}, {Inoue}, {Koyama}, {Makishima},
  {Matsuoka}, {Ogawara}, {Shibazaki}, {Suzuki}, {Tanaka}, \&
  {Hirano}}]{1984PASJ...36..741M}
{Mitsuda}, K., {Inoue}, H., {Koyama}, K., {et~al.} 1984, \pasj, 36, 741

\bibitem[{{Parker} {et~al.}(2016){Parker}, {Tomsick}, {Kennea}, {Miller},
  {Harrison}, {Barret}, {Boggs}, {Christensen}, {Craig}, {Fabian}, {F{\"u}rst},
  {Grinberg}, {Hailey}, {Romano}, {Stern}, {Walton}, \&
  {Zhang}}]{2016ApJ...821L...6P}
{Parker}, M.~L., {Tomsick}, J.~A., {Kennea}, J.~A., {et~al.} 2016, \apjl, 821,
  L6, \dodoi{10.3847/2041-8205/821/1/L6}

\bibitem[{{Plant} {et~al.}(2014){Plant}, {Fender}, {Ponti}, {Mu{\~n}oz-Darias},
  \& {Coriat}}]{2014MNRAS.442.1767P}
{Plant}, D.~S., {Fender}, R.~P., {Ponti}, G., {Mu{\~n}oz-Darias}, T., \&
  {Coriat}, M. 2014, \mnras, 442, 1767, \dodoi{10.1093/mnras/stu867}

\bibitem[{{Plant} {et~al.}(2015){Plant}, {Fender}, {Ponti}, {Mu{\~n}oz-Darias},
  \& {Coriat}}]{2015A&A...573A.120P}
---. 2015, \aap, 573, A120, \dodoi{10.1051/0004-6361/201423925}

\bibitem[{{Poutanen} {et~al.}(2018){Poutanen}, {Veledina}, \&
  {Zdziarski}}]{2018A&A...614A..79P}
{Poutanen}, J., {Veledina}, A., \& {Zdziarski}, A.~A. 2018, \aap, 614, A79,
  \dodoi{10.1051/0004-6361/201732345}

\bibitem[{{Reis} {et~al.}(2008){Reis}, {Fabian}, {Ross}, {Miniutti}, {Miller},
  \& {Reynolds}}]{2008MNRAS.387.1489R}
{Reis}, R.~C., {Fabian}, A.~C., {Ross}, R.~R., {et~al.} 2008, \mnras, 387,
  1489, \dodoi{10.1111/j.1365-2966.2008.13358.x}

\bibitem[{{Shakura} \& {Sunyaev}(1973)}]{1973A&A....24..337S}
{Shakura}, N.~I., \& {Sunyaev}, R.~A. 1973, \aap, 24, 337

\bibitem[{{Shaposhnikov} \& {Titarchuk}(2009)}]{2009ApJ...699..453S}
{Shaposhnikov}, N., \& {Titarchuk}, L. 2009, \apj, 699, 453,
  \dodoi{10.1088/0004-637X/699/1/453}

\bibitem[{{Singh} {et~al.}(2016){Singh}, {Stewart}, {Chandra}, {Mukerjee},
  {Kotak}, {Beardmore}, {Chitnis}, {Dewangan}, {Bhattacharyya}, {Mirza},
  {Kamble}, {Navalkar}, {Shah}, {Vishwakarma}, \&
  {Koyande}}]{2016SPIE.9905E..1ES}
{Singh}, K.~P., {Stewart}, G.~C., {Chandra}, S., {et~al.} 2016, in Society of
  Photo-Optical Instrumentation Engineers (SPIE) Conference Series, Vol. 9905,
  Space Telescopes and Instrumentation 2016: Ultraviolet to Gamma Ray, ed.
  J.-W.~A. {den Herder}, T.~{Takahashi}, \& M.~{Bautz}, 99051E,
  \dodoi{10.1117/12.2235309}

\bibitem[{{Singh} {et~al.}(2017){Singh}, {Stewart}, {Westergaard},
  {Bhattacharayya}, {Chandra}, {Chitnis}, {Dewangan}, {Kothare}, {Mirza},
  {Mukerjee}, {Navalkar}, {Shah}, {Abbey}, {Beardmore}, {Kotak}, {Kamble},
  {Vishwakarama}, {Pathare}, {Risbud}, {Koyande}, {Stevenson}, {Bicknell},
  {Crawford}, {Hansford}, {Peters}, {Sykes}, {Agarwal}, {Sebastian},
  {Rajarajan}, {Nagesh}, {Narendra}, {Ramesh}, {Rai}, {Navalgund}, {Sarma},
  {Pandiyan}, {Subbarao}, {Gupta}, {Thakkar}, {Singh}, \&
  {Bajpai}}]{2017JApA...38...29S}
{Singh}, K.~P., {Stewart}, G.~C., {Westergaard}, N.~J., {et~al.} 2017, Journal
  of Astrophysics and Astronomy, 38, 29, \dodoi{10.1007/s12036-017-9448-7}

\bibitem[{{Sreehari} {et~al.}(2019){Sreehari}, {Iyer}, {Radhika}, {Nandi}, \&
  {Mandal}}]{2019AdSpR..63.1374S}
{Sreehari}, H., {Iyer}, N., {Radhika}, D., {Nandi}, A., \& {Mandal}, S. 2019,
  Advances in Space Research, 63, 1374, \dodoi{10.1016/j.asr.2018.10.042}

\bibitem[{{Sridhar} {et~al.}(2019){Sridhar}, {Bhattacharyya}, {Chandra}, \&
  {Antia}}]{2019MNRAS.487.4221S}
{Sridhar}, N., {Bhattacharyya}, S., {Chandra}, S., \& {Antia}, H.~M. 2019,
  \mnras, 487, 4221, \dodoi{10.1093/mnras/stz1476}

\bibitem[{{van der Klis}(1995)}]{1995xrbi.nasa..252V}
{van der Klis}, M. 1995, in X-ray Binaries, 252--307

\bibitem[{{van der Klis}(2006)}]{2006csxs.book...39V}
{van der Klis}, M. 2006, in Compact stellar X-ray sources, Vol.~39, 39--112

\bibitem[{{Verner} {et~al.}(1996){Verner}, {Ferland}, {Korista}, \&
  {Yakovlev}}]{1996ApJ...465..487V}
{Verner}, D.~A., {Ferland}, G.~J., {Korista}, K.~T., \& {Yakovlev}, D.~G. 1996,
  \apj, 465, 487, \dodoi{10.1086/177435}

\bibitem[{{Wardzi{\'n}ski} {et~al.}(2002){Wardzi{\'n}ski}, {Zdziarski},
  {Gierli{\'n}ski}, {Grove}, {Jahoda}, \& {Johnson}}]{2002MNRAS.337..829W}
{Wardzi{\'n}ski}, G., {Zdziarski}, A.~A., {Gierli{\'n}ski}, M., {et~al.} 2002,
  \mnras, 337, 829, \dodoi{10.1046/j.1365-8711.2002.05914.x}

\bibitem[{{Wilms} {et~al.}(2000){Wilms}, {Allen}, \&
  {McCray}}]{2000ApJ...542..914W}
{Wilms}, J., {Allen}, A., \& {McCray}, R. 2000, \apj, 542, 914,
  \dodoi{10.1086/317016}

\bibitem[{{Yadav} {et~al.}(2016{\natexlab{a}}){Yadav}, {Agrawal}, {Antia},
  {Chauhan}, {Dedhia}, {Katoch}, {Madhwani}, {Manchanda}, {Misra}, {Pahari},
  {Paul}, \& {Shah}}]{2016SPIE.9905E..1DY}
{Yadav}, J.~S., {Agrawal}, P.~C., {Antia}, H.~M., {et~al.} 2016{\natexlab{a}},
  in Society of Photo-Optical Instrumentation Engineers (SPIE) Conference
  Series, Vol. 9905, Space Telescopes and Instrumentation 2016: Ultraviolet to
  Gamma Ray, ed. J.-W.~A. {den Herder}, T.~{Takahashi}, \& M.~{Bautz}, 99051D,
  \dodoi{10.1117/12.2231857}

\bibitem[{{Yadav} {et~al.}(2016{\natexlab{b}}){Yadav}, {Misra}, {Verdhan
  Chauhan}, {Agrawal}, {Antia}, {Pahari}, {Dedhia}, {Katoch}, {Madhwani},
  {Manchanda}, {Paul}, {Shah}, \& {Ishwara-Chandra}}]{2016ApJ...833...27Y}
{Yadav}, J.~S., {Misra}, R., {Verdhan Chauhan}, J., {et~al.}
  2016{\natexlab{b}}, \apj, 833, 27, \dodoi{10.3847/0004-637X/833/1/27}

\bibitem[{{Yamada} {et~al.}(2009){Yamada}, {Makishima}, {Uehara}, {Nakazawa},
  {Takahashi}, {Dotani}, {Ueda}, {Ebisawa}, {Kubota}, \&
  {Gandhi}}]{2009ApJ...707L.109Y}
{Yamada}, S., {Makishima}, K., {Uehara}, Y., {et~al.} 2009, \apjl, 707, L109,
  \dodoi{10.1088/0004-637X/707/2/L109}

\bibitem[{{Yang} {et~al.}(2023){Yang}, {Zhang}, {Zhang}, {M{\'e}ndez},
  {Garc{\'\i}a}, {Huang}, {Bu}, {Liu}, {Yu}, {Wang}, {Tao}, {Altamirano}, {Qu},
  {Zhang}, {Ma}, {Song}, {Jia}, {Ge}, {Liu}, {Yan}, {Li}, {Ren}, {Ma}, {Zhang},
  {Xu}, {Ma}, {Du}, {Fu}, {Xiao}, {Li}, {Jin}, {Zhao}, \&
  {Zhao}}]{2023MNRAS.521.3570Y}
{Yang}, Z.-X., {Zhang}, L., {Zhang}, S.~N., {et~al.} 2023, \mnras, 521, 3570,
  \dodoi{10.1093/mnras/stad795}

\bibitem[{{Zdziarski} \& {De Marco}(2020)}]{2020ApJ...896L..36Z}
{Zdziarski}, A.~A., \& {De Marco}, B. 2020, \apjl, 896, L36,
  \dodoi{10.3847/2041-8213/ab9899}

\bibitem[{{Zdziarski} {et~al.}(2021{\natexlab{a}}){Zdziarski}, {De Marco},
  {Szanecki}, {Nied{\'z}wiecki}, \& {Markowitz}}]{2021ApJ...906...69Z}
{Zdziarski}, A.~A., {De Marco}, B., {Szanecki}, M., {Nied{\'z}wiecki}, A., \&
  {Markowitz}, A. 2021{\natexlab{a}}, \apj, 906, 69,
  \dodoi{10.3847/1538-4357/abca9c}

\bibitem[{{Zdziarski} {et~al.}(2021{\natexlab{b}}){Zdziarski}, {Dzie{\l}ak},
  {De Marco}, {Szanecki}, \& {Nied{\'z}wiecki}}]{2021ApJ...909L...9Z}
{Zdziarski}, A.~A., {Dzie{\l}ak}, M.~A., {De Marco}, B., {Szanecki}, M., \&
  {Nied{\'z}wiecki}, A. 2021{\natexlab{b}}, \apjl, 909, L9,
  \dodoi{10.3847/2041-8213/abe7ef}

\bibitem[{{Zdziarski} {et~al.}(2004){Zdziarski}, {Gierli{\'n}ski},
  {Miko{\l}ajewska}, {Wardzi{\'n}ski}, {Smith}, {Harmon}, \&
  {Kitamoto}}]{2004MNRAS.351..791Z}
{Zdziarski}, A.~A., {Gierli{\'n}ski}, M., {Miko{\l}ajewska}, J., {et~al.} 2004,
  \mnras, 351, 791, \dodoi{10.1111/j.1365-2966.2004.07830.x}

\bibitem[{{Zdziarski} {et~al.}(1996){Zdziarski}, {Johnson}, \&
  {Magdziarz}}]{1996MNRAS.283..193Z}
{Zdziarski}, A.~A., {Johnson}, W.~N., \& {Magdziarz}, P. 1996, \mnras, 283,
  193, \dodoi{10.1093/mnras/283.1.193}

\bibitem[{{Zdziarski} {et~al.}(1998){Zdziarski}, {Poutanen}, {Mikolajewska},
  {Gierlinski}, {Ebisawa}, \& {Johnson}}]{1998MNRAS.301..435Z}
{Zdziarski}, A.~A., {Poutanen}, J., {Mikolajewska}, J., {et~al.} 1998, \mnras,
  301, 435, \dodoi{10.1046/j.1365-8711.1998.02021.x}

\bibitem[{{Zdziarski} {et~al.}(2022){Zdziarski}, {You}, {Szanecki}, {Li}, \&
  {Ge}}]{2022ApJ...928...11Z}
{Zdziarski}, A.~A., {You}, B., {Szanecki}, M., {Li}, X.-B., \& {Ge}, M. 2022,
  \apj, 928, 11, \dodoi{10.3847/1538-4357/ac54a7}

\bibitem[{{Zdziarski} {et~al.}(2019){Zdziarski}, {Zi{\'o}{\l}kowski}, \&
  {Miko{\l}ajewska}}]{2019MNRAS.488.1026Z}
{Zdziarski}, A.~A., {Zi{\'o}{\l}kowski}, J., \& {Miko{\l}ajewska}, J. 2019,
  \mnras, 488, 1026, \dodoi{10.1093/mnras/stz1787}

\bibitem[{{{\.Z}ycki} {et~al.}(1999){{\.Z}ycki}, {Done}, \&
  {Smith}}]{1999MNRAS.309..561Z}
{{\.Z}ycki}, P.~T., {Done}, C., \& {Smith}, D.~A. 1999, \mnras, 309, 561,
  \dodoi{10.1046/j.1365-8711.1999.02885.x}

\end{thebibliography}
\bibliographystyle{aasjournal}



\end{document}